# Thermally modulated cross-stream migration of a surfactant-laden deformable drop in a Poiseuille flow

Sayan Das and Suman Chakraborty[a]

*Department of Mechanical Engineering, Indian Institute of Technology Kharagpur, West Bengal - 721302, India*

In the present study, we investigate the cross-stream migration of a deformable droplet suspended in a non-isothermal Poiseuille flow in the presence of bulk-insoluble surfactants. Owing to the non-linearity present in the system of governing equations, an asymptotic approach is adopted, in an effort to capture the intricate and non-trivial coupling between the various influencing parameters. With the assumption of negligible inertia in fluid flow and convective transport of thermal energy, we obtain the droplet migration velocity through small-deformation perturbation analysis for two different limiting cases, namely, convection-driven-surfactant transport and surface-diffusion-dominated surfactant transport. Under each of these limiting cases, the cross-stream migration of droplet is studied for a constant temperature gradient applied in the same direction as well as in a direction opposite to the imposed flow. For the former limiting case, the droplet is always migrates towards the centerline of flow. For a highly viscous droplet, the direction of its cross-stream migration reverses. When the temperature decreases in the direction of the imposed flow, cross-stream migration velocity reduces with increase in the applied temperature gradient till a critical point is reached at which there is no cross-stream migration. Beyond the critical point, there is a gradual increase in the magnitude of the cross-stream velocity. The droplet, below the critical temperature gradient, migrates towards the flow centerline; however, above it the droplet moves away from the centerline. For the other limiting case of surfactant transport dominated by surface convection, the magnitude of the cross-stream velocity is found to be significantly larger and at the same time independent of the droplet-carrier phase viscosity ratio.

[a] E-mail address for correspondence: suman@mech.iitkgp.ernet.in

# I. INTRODUCTION

Study on droplet dynamics in pressure driven flows is a rising area of research due to its wide spread applications in different microfluidic devices [1–3]. A significant number of such applications are directed towards medical diagnostics and material processing industries, including specific examples in drug delivery, cell encapsulation, reagent mixing and analytic detection [1,4–7]. Some other relevant biological applications include the transverse migration and positioning of red blood cells and erythrocytes in the flow of blood through arteries [8,9]. A proper methodology for modulating the position of the dispersed phase (cells, droplets or particles) has a wide scope in the domain of flow fractionation [10,11] and cytometry [12]. This control over the steady-state lateral and axial position of the droplet can be fine-tuned with the aid of an externally imposed temperature gradient, which can also be used for separation and sorting of droplets [13–15].

Suspended droplets are transported with the help of syringe pumps in a wide variety of microfluidic devices [16]. Several theoretical as well as experimental studies have been executed to analyze the dynamics of droplets under such situations [16–18]. Haber and Hetsroni [19] theoretically demonstrated the migration characteristics of a surfactant-free, non-deformable and a Newtonian droplet suspended in an isothermal arbitrary Stokes flow. Later, the effect of different non-linear effects such as shape deformation, inertia, viscoelasticity were investigated [18,20,21]. It was shown that a deformable droplet, when placed eccentrically with respect to the centerline of a pressure driven, exhibits a migration in the cross-stream direction [20,22–26]. Chan and Leal [20] showed the effect of $\lambda$ (which is the ratio of viscosity of the droplet phase to the carrier phase) on the cross-stream migration of the droplet. They showed that the same droplet which had a cross-stream migration towards the flow centerline for $\lambda < 0.5$ and $\lambda > 10$, moves away from the centerline for $0.5 < \lambda < 10$. The cross-stream migration of the droplet is affected further under the presence of different external effects such as viscoelasticity [20,27,28], fluid inertia [18,29,30]. In some of the recently published works, it was shown that cross-stream migration of a droplet exists even though there was no deformation, inertia or viscoelasticity involved. This was due to the presence of surfactants [31,32] and an imposed temperature gradient [33,34].

The presence of surfactants or contaminants in a microfluidic device is quite common. Non-uniformly distributed surfactants along the droplet surface bring about significant alteration to its dynamics by altering the surface tension along its surface [32,35]. There is experimental evidence which proves that there exists a relationship between the shape deformation and surfactant distribution along the droplet surface [36–38]. Stan et al. [16] showed numerically as well as experimentally that shape deformation of a surfactant-free droplet has a important role in altering the droplet dynamics. Deformation of droplet induces a lift force that significantly affects the cross-stream migration of the droplet. However, a significant number of studies have also analyzed the migration characteristics of a non-deformable surfactant laden droplet [32,33]. The presence of a non-uniform distribution of surfactants generates a Marangoni stress along the interface of the droplet. Vlahovska et al. [39] used a small-deformation perturbation approach to show the effect of this Marangoni stress on the dynamics of a droplet suspended in a linear flow, provided the surfactant transport is dominated by surface convection.

A significant amount of research is directed towards proper modulation of the droplet or a particle in a flow field due to presence of different external effects such as magnetic [2], acoustic [2], electric [40–42] or temperature [43]. The present study, however, takes into account the role of temperature field in altering the droplet dynamics. Variation of temperature in the flow field and hence along the droplet surface changes the interfacial tension. This generates a Marangoni stress which causes an imbalance in the stress balance and hence alters the net force acting on the droplet affecting droplet migration. Young et al. [44] were the first to theoretically obtain the axial droplet migration velocity in a linearly varying temperature flow field without the presence of any imposed flow or any nonlinearities such as shape deformation or inertial effects. Later a number of researchers have studied thermocapillary motion of droplets under the premise several aspects such as fluid inertia [45], droplet deformation [46], thermal convection [47–49] and bounding walls [50–54]. Sekhar et al. [55], in a recent work studied the thermocapillary migration of droplet in the presence of an imposed Stokes flow. Their study shows that the effects due to imposed flow and thermocapillary action on fluid flow can be linearly combined in the absence of any shape deformation.

A number of studies have also shown the combined effect of thermocapillary and surfactant-induced Marangoni stress on droplet migration characteristics [33,56]. In a recent work, Das et al. [34] showed the combined effect of temperature variation and surfactant distribution on the dynamics of a non-deformable droplet suspended in a Poiseuille flow. However, there is no such study available in the literature that focuses on the migration characteristics of a deformable droplet suspended in a Poiseuille flow under the combined presence of a bulk insoluble surfactants and a constant temperature gradient. In this present study, we study the effect of temperature distribution, shape deformation and associated surfactant redistribution on the cross-stream migration of the droplet. This problem is nonlinear and non-trivial, and cannot be addressed as a mere linear superposition of the concerned influencing parameters, due to the consideration of shape deformation. As a consequence, the associated governing differential equations are coupled due to the convection of surfactants along the droplet surface. Hence, a linear superposition of the results of a thermocapillary-driven and an imposed flow actuated migration of a surfactant-laden droplet may give erroneous predictions. In order to tackle such a situation, we use an asymptotic approach for two limiting cases, namely, surface convection-dominated-surfactant transport and surface diffusion-dominated-surfactant transport. How a change in the imposed temperature gradient and shape deformation-induced surfactant redistribution alters both the magnitude as well as the direction of the cross-stream migration velocity of the droplet is the prime objective of this study.

## II. PROBLEM FORMULATION

### A. System description

The present system consists of a neutrally buoyant Newtonian droplet of radius $a$ suspended in another Newtonian fluid with an imposed Poiseuille flow. A schematic of the system is given in Fig. 1. The thermal conductivity and bulk viscosity of the droplet phase is denoted by $k_i$ and $\eta_i$ respectively. The respective properties for the suspending phase are $k_e$ and $\eta_e$. It is assumed that a constant temperature gradient, $\bar{G}$, is applied to the suspending phase. Bulk insoluble surfactants are assumed to be present on the interface of the droplet. The surfactants gets transported only along the surface of the droplet by means of convection as well as diffusion. The only property which is not constant is the surface tension $(\bar{\sigma})$. The surface tension is directly dependent on the interfacial temperature $\bar{T}_s$ and the local surfactant concentration $(\bar{\Gamma})$. The equilibrium surface tension for a surfactant-free droplet is denoted by $\bar{\sigma}_c$. In the absence of any imposed flow or temperature gradient, the surfactant is uniformly distributed along the droplet surface. The equilibrium surfactant concentration under such a situation is represented by $\bar{\Gamma}_{eq}$. The corresponding equilibrium surface tension is denoted by $\bar{\sigma}_{eq}$ that eventually changes to $\bar{\sigma}$ due to the disturbance generated by the imposed flow and a constant temperature gradient. The variation of temperature at the droplet interface induces a Marangoni stress which further causes transport of the surfactants along the droplet surface. Thus the combined presence of an imposed Poiseuille flow and a linearly varying temperature field generates Marangoni stresses which alters the migration velocity of the droplet $(\bar{\mathbf{U}})$. The primary objective of the present study is to analyze the effect of the thermal Marangoni stresses on the cross-stream migration characteristics of the droplet. As can be seen from Fig. 1, we use a spherical coordinate system $(\bar{r}, \theta, \varphi)$ which is attached to the centroid of the droplet.

### B. Important assumptions

The problem at hand can be solved analytically only if certain assumptions are made. Some of the important assumptions required to simplify the governing equations and boundary conditions for flow and temperature field are as follows

(i) The transport of thermal energy by means of advection is neglected. In other words the conduction is considered to be the main mode of energy transport. Thus the thermal Péclet number for the present study is taken to be small, $Pe_T = \bar{V}_c a / \alpha_e \ll 1$, where $\alpha_e$ is the thermal diffusivity of the continuous phase and $\bar{V}_c$ is the centerline velocity of the imposed Poiseuille flow.

(ii) The fluid flow is assumed to be governed by viscous, pressure and surface tension forces, that is, the effect of fluid inertia is neglected. Thus the flow Reynolds number based on droplet radius is taken to be small $Re = \rho \bar{V}_c a / \mu_e \ll 1$, where $\rho$ is the density of either of the phases.

(iii) Only small deformation of the droplet is taken into account. The surface tension forces prevents the viscous forces from deforming the droplet largely from its original spherical shape. Thus the capillary number, $Ca^* = \mu_e \bar{V}_c / \bar{\sigma}_{eq}$, which is the ratio of the viscous force to the surface tension force acting on the droplet is assumed to be small $(Ca^* \ll 1)$.

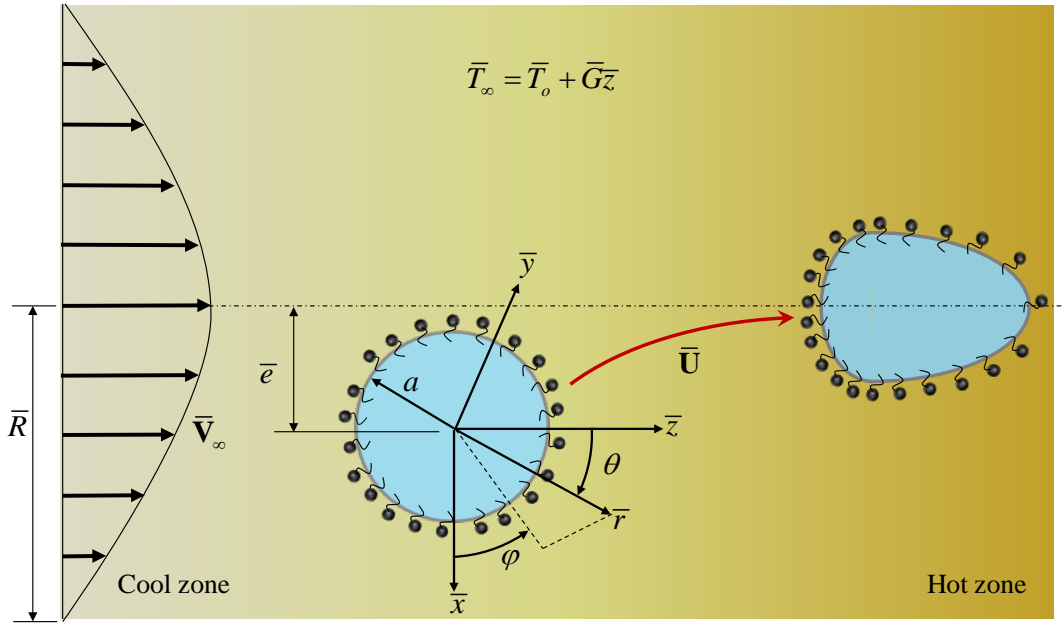

Fig. 1. Schematic of a surfactant laden droplet of radius $a$ suspended in a Poiseuille flow field. A linearly increasing temperature field $(\bar{T}_\infty)$ in the direction of the imposed flow $(\bar{z}\ \text{direction})$ is shown in the Fig.. The droplet is placed at an off center position, $\bar{e}$ being the distance between the centroid of the droplet and center line of the flow. $\bar{R}$ is the distance from centerline of flow to the point of zero velocity. Both spherical $(\bar{r}, \theta, \varphi)$ as well as cartesian coordinate system $(\bar{x}, \bar{y}, \bar{z})$ are shown. The $\bar{x}$ axis is directed away from the centerline of the droplet.

(iv) The surfactants only gets transported along the droplet surface and is insoluble in either of the phases. [57]

(v) The local surfactant concentration along the droplet surface is independent of the temperature distribution along the surface of the droplet and hence it is not affected by the heat transfer process. [56]

(vi) The interfacial tension is linearly dependent on the surfactant concentration and the interfacial temperature distribution along the surface of the droplet through the equation of state. [58,59]

(vii) Any effect of bounding walls is neglected, that is the droplet is assumed to be suspended in an unbounded medium.

### C. Experimental Relevance

Typical values of the above non-dimensional numbers can be obtained from the experimental work of Chen et. al [60], where they investigated the thermocapillary migration of a surfactant laden droplet. They used a neutrally-buoyant system comprised of water as the dispersed phase and n-butyl benzoate (with $\rho_e = 995 \text{ kg/m}^3$, $\mu_e = 2.49 \times 10^{-3} \text{ Ns/m}^2$, $\alpha_e = 6.63 \times 10^{-8} \text{ m}^2/\text{s}$,) as the continuous phase. We take into consideration the diameter of the droplet as $a = 50\,\mu\text{m}$, the characteristic velocity of fluid flow as $\bar{V}_C = 10^{-4} \text{ m/s}$ and a characteristic temperature difference of $\Delta T = 2\,°\text{C}$ for the above system. The equilibrium surface tension at a water-n-butyl benzoate interface is $\bar{\sigma}_{eq} = 22.76 \times 10^{-3} \text{ N/m}$ at a reference temperature of $\bar{T}_o = 25\,°\text{C}$. With the use of the above property values as used in the experimental work of Chen et al. [60], we obtain the different non-dimensional numbers defined above as: $Re = 0.002$, $Pe_T = 7.54 \times 10^{-2}$, $Ca^* = 0.109 \times 10^{-4}$. These values are in direct agreement with the assumptions made above.

### D. Governing equations and boundary conditions

The governing equation for the temperature field is the energy equation. However, under the assumption of low thermal Péclet number, the energy equation reduces to a Laplace equation of the following form

$$\left.\begin{aligned}\bar{\nabla}^2 \bar{T}_i &= 0, \\ \bar{\nabla}^2 \bar{T}_e &= 0,\end{aligned}\right\} \quad (1)$$

where $\bar{T}_i$ and $\bar{T}_e$ represent the temperature inside and outside the droplet, respectively. The quantities with subscript '$i$' represent the droplet phase, while those with subscript '$e$' represent the carrier phase. The far field condition for the temperature field is satisfied by the temperature outside the droplet, and is given by

$$\text{as } \bar{r} \to \infty, \ \bar{T}_e = \bar{T}_\infty, \quad (2)$$

where $\bar{T}_\infty$ represents the temperature at the far field and can be expressed as follows

$$\bar{T}_\infty = \bar{T}_o + \bar{G}\bar{z}, \tag{3}$$

where, $\bar{T}_o$ is the reference temperature. This temperature field is applied in the same direction as that of the imposed Poiseuille flow. Depending on whether $\bar{G} > 0$ or $\bar{G} < 0$ the temperature in the far field increases or decreases in the direction of the Poiseuille flow, respectively. The temperature inside the droplet $(\bar{T}_i)$ is bounded at the center, $\bar{r} = 0$. The other boundary conditions at the droplet interface $(\bar{r} = \bar{r}_s)$ include the continuity in temperature and heat transfer, which can be expressed as

$$\left. \begin{array}{l} \text{at } \bar{r} = \bar{r}_s, \quad \bar{T}_i = \bar{T}_e, \\ \text{at } \bar{r} = \bar{r}_s, \quad k_i \left( \bar{\nabla} \bar{T}_i \right) \cdot \hat{\mathbf{n}} = k_e \left( \bar{\nabla} \bar{T}_e \right) \cdot \hat{\mathbf{n}}. \end{array} \right\} \tag{4}$$

The flow field is governed by the continuity and the Stokes equation, as the inertia terms are neglected on assumption of low Reynolds number flow. The governing equations for the flow field are thus represented as

$$\left. \begin{array}{l} -\bar{\nabla}\bar{p}_i + \mu_i \bar{\nabla}^2 \bar{\mathbf{u}}_i = \mathbf{0}, \quad \bar{\nabla} \cdot \bar{\mathbf{u}}_i = 0, \\ -\bar{\nabla}\bar{p}_e + \mu_e \bar{\nabla}^2 \bar{\mathbf{u}}_e = \mathbf{0}, \quad \bar{\nabla} \cdot \bar{\mathbf{u}}_e = 0, \end{array} \right\} \tag{5}$$

where, $(\bar{\mathbf{u}}, \bar{p})$ are the velocity and pressure fields. The velocity and pressure fields outside the droplet satisfy the far-field condition which is given by

$$\left. \begin{array}{l} \text{as } \bar{r} \to \infty, \quad \bar{\mathbf{u}}_e = \bar{\mathbf{V}}_\infty - \bar{\mathbf{U}}, \\ \text{as } \bar{r} \to \infty, \quad \bar{p}_e = \bar{p}_\infty, \end{array} \right. \tag{6}$$

where $\bar{p}_\infty$ is the far-field pressure corresponding to $\bar{\mathbf{V}}_\infty$, which is the imposed Poiseuille flow field. $\bar{\mathbf{V}}_\infty$ can be expressed in terms of the spherical coordinates (attached to the center of the droplet) in the following manner

$$\bar{\mathbf{V}}_\infty = \bar{V}_c \left[ 1 - \frac{\bar{e}^2}{\bar{R}^2} - \frac{\bar{r}^2}{\bar{R}^2} \sin^2 \theta - \frac{2\bar{r}}{\bar{R}^2} \bar{e} \cos\varphi \sin\theta \right] \mathbf{e}_z, \tag{7}$$

where $\bar{e}$ represents the eccentricity of the droplet and $\bar{R}$ is the distance from the centerline of flow to the point of zero velocity. Both the velocity and pressure fields inside the droplet, $(\bar{\mathbf{u}}_i, \bar{p}_i)$ are bounded at the center, $\bar{r} = 0$. At the interface $(\bar{r} = \bar{r}_s)$, the velocity and pressure fields satisfy the following boundary conditions:

$$\begin{aligned}
&\text{at } \bar{r} = \bar{r}_s, \quad \bar{\mathbf{u}}_i = \bar{\mathbf{u}}_e, \\
&\text{at } \bar{r} = \bar{r}_s, \quad \bar{\mathbf{u}}_i \cdot \hat{\mathbf{n}} = \bar{\mathbf{u}}_e \cdot \hat{\mathbf{n}} = 0, \\
&\text{at } \bar{r} = \bar{r}_s, \quad \left(\bar{\boldsymbol{\tau}}_e \cdot \hat{\mathbf{n}} - \bar{\boldsymbol{\tau}}_i \cdot \hat{\mathbf{n}}\right) = -\bar{\nabla}_s \bar{\sigma} + \bar{\sigma}\left(\bar{\nabla} \cdot \hat{\mathbf{n}}\right)\hat{\mathbf{n}},
\end{aligned} \qquad (8)$$

where, $\bar{\boldsymbol{\tau}}_i = -\bar{p}_i \mathbf{I} + \mu_i \left[\bar{\nabla}\bar{\mathbf{u}}_i + \left(\bar{\nabla}\bar{\mathbf{u}}_i\right)^T\right]$ and $\bar{\boldsymbol{\tau}}_e = -\bar{p}_e \mathbf{I} + \mu_e \left[\bar{\nabla}\bar{\mathbf{u}}_e + \left(\bar{\nabla}\bar{\mathbf{u}}_e\right)^T\right]$ are respectively the hydrodynamic stresses tensors inside and outside the droplet comprising of a hydrostatic and a deviatoric component. $\bar{\nabla}_s = \left[\bar{\nabla}(\mathbf{I} - \mathbf{nn})\right]$ in the above equation represents the surface gradient operator. First boundary condition in the above set of equations represents the continuity of velocity field across the interface, the second boundary condition represents the kinematic boundary condition or the no-penetration boundary condition and finally the last boundary condition is the stress balance condition. $\hat{\mathbf{n}}$ is a unit normal drawn perpendicular to the surface of the droplet and is given by

$$\mathbf{n} = \frac{\bar{\nabla} F}{\left|\bar{\nabla} F\right|}, \qquad (9)$$

where, $\bar{F} = \bar{r} - \bar{r}_s$ represents the surface of the droplet.

The temperature induced- as well as the surfactant induced-Marangoni stress both depend on the variation of the surface tension $(\bar{\sigma})$ with the temperature and surfactant concentration on the surface of the droplet, respectively. As per our assumption we assume a linear relationship between the interfacial surface tension with the surfactant concentration $(\bar{\Gamma})$ and the temperature distribution $(\bar{T}_s)$ along the droplet surface. This can be written as [33,56]

$$\bar{\sigma} = \bar{\sigma}_c - \gamma_T \left(\bar{T}_s - \bar{T}_o\right) - R_g \bar{T}_o \bar{\Gamma}, \qquad (10)$$

where, $\bar{T}_s \left(= \bar{T}\big|_{\bar{r}=\bar{r}_s}\right)$ is the temperature at the droplet interface, $\gamma_T = -d\bar{\sigma}/d\bar{T}_s$ represents the rate of variation of surface tension with the surface temperature and $R_g$ is the universal gas constant.

The local surfactant concentration is governed by a convection-diffusion equation along the surface of the droplet, which is given by [33,56]

$$\bar{\nabla}_s \cdot \left(\bar{\mathbf{u}}_s \bar{\Gamma}\right) = D_s \bar{\nabla}_s^2 \bar{\Gamma}, \qquad (11)$$

where, $\bar{\mathbf{u}}_s$ is the surface velocity and $D_s$ is the surface diffusivity of the surfactant.

### E. Dimensionless form of governing equations and boundary conditions

Now that we have stated the governing differential equations and boundary conditions in its dimensional form, we derive the non-dimensional version of the same. Towards this, we first use the following non-dimensional scheme

$$\left. \begin{array}{l} r = \bar{r}/a,\, \mathbf{u} = \bar{\mathbf{u}}/\bar{V}_c,\, \Gamma = \bar{\Gamma}/\bar{\Gamma}_{eq},\, \sigma = \bar{\sigma}/\bar{\sigma}_c, \\ p = \bar{p}/(\mu_e \bar{V}_c/a),\, \boldsymbol{\tau} = \bar{\boldsymbol{\tau}}/(\mu_e \bar{V}_c/a),\, T = (\bar{T} - \bar{T}_o)/|\bar{G}|a, \end{array} \right\} \quad (12)$$

where, all the quantities with an 'overbar' represents dimensional quantities and those without any 'overbar' are dimensionless quantities. While deriving the governing equations and boundary conditions we encounter various non-dimensional entities such as: (i) the viscosity ratio, $\lambda = \mu_i/\mu_e$, which is the ratio of the viscosity of the droplet phase to that of the suspending phase; (ii) the thermal conductivity ratio, $\delta = k_i/k_e$, which is the ratio of the thermal conductivity of the droplet phase to that of the carrier phase; (iii) the elasticity number, $\beta = \bar{\Gamma}_{eq} RT_o/\bar{\sigma}_c = -d(\bar{\sigma}/\bar{\sigma}_c)/d\bar{\Gamma}$, which indicates the sensitivity of surface tension to the local surfactant concentration on the surface of the droplet; (iv) the thermal Marangoni number, $Ma_T = \gamma_T |\bar{G}| a/\mu_e \bar{V}_c$, which is the ratio of the thermocapillary induced Marangoni stress to the viscous stress ; (v) the surface Péclet number, $Pe_s = \bar{V}_c a/D_s$, which signifies the relative importance of surfactant transport due to advection to that due to surface diffusion; (vi) the modified Capillary number, $Ca = Ca^*/(1-\beta)$, which is the ratio of the viscous force to the surface tension force acting on the droplet. From the definition of $\beta$, the equilibrium surface tension for a surfactant laden droplet in the absence of any imposed flow or temperature gradient can be expressed as $\bar{\sigma}_{eq} = \bar{\sigma}_c (1-\beta)$. Since, in our analysis we deal with a surfactant-laden droplet, it is more convenient to work with $\bar{\sigma}_{eq}$ instead of $\bar{\sigma}_c$. This is the sole reason for defining a modified Capillary number, $Ca$.

With the help above mentioned scaling parameters and the non-dimensional numbers, we non-dimensionalise the governing equations as well as the relevant boundary conditions. The dimensionless governing equations for the thermal problem can be written as

$$\left. \begin{array}{l} \nabla^2 T_i = 0, \\ \nabla^2 T_e = 0, \end{array} \right\} \quad (13)$$

subjected to the following boundary conditions at the interface

$$\left.\begin{aligned}&\text{as } r \to \infty, T_e = \zeta r \cos\theta, \\ & T_i \text{ is bounded at } r = 0, \\ & \text{at } r = r_s, \quad T_i = T_e, \\ & \text{at } r = r_s, \quad \delta(\nabla T_i)\cdot\hat{\mathbf{n}} = (\nabla T_e)\cdot\hat{\mathbf{n}}.\end{aligned}\right\} \quad (14)$$

The quantity $\zeta$ indicates whether the droplet migrates in the direction of the imposed flow field or against it. If $\zeta = 1$, the temperature increases in the direction of the imposed flow (positive z direction), whereas $\zeta = -1$ signifies that the temperature decreases in the direction opposite to the direction of the imposed flow (negative z direction). The governing differential equations for the flow field can be expressed in a non-dimensional form as follows [61]

$$\left.\begin{aligned}&-\nabla p_i + \lambda \nabla^2 \mathbf{u}_i = \mathbf{0}, \ \nabla \cdot \mathbf{u}_i = 0, \\ & -\nabla p_e + \nabla^2 \mathbf{u}_e = \mathbf{0}, \ \nabla \cdot \mathbf{u}_e = 0.\end{aligned}\right\} \quad (15)$$

The relevant boundary conditions at the far-field and at the interface of the droplet are mentioned below

$$\left.\begin{aligned}&\text{at } r \to \infty, \ (\mathbf{u}_e, p_e) = (\mathbf{V}_\infty - \mathbf{U}, p_\infty), \\ & \mathbf{u}_i \text{ is bounded at } r = 0, \\ & \text{at } r = r_s, \ \mathbf{u}_i \cdot \hat{\mathbf{n}} = \mathbf{u}_e \cdot \hat{\mathbf{n}} = 0, \\ & \text{at } r = r_s, \ \mathbf{u}_i = \mathbf{u}_e, \\ & \text{at } r = r_s, \ (\boldsymbol{\tau}_e \cdot \hat{\mathbf{n}} - \boldsymbol{\tau}_i \cdot \hat{\mathbf{n}}) = \zeta Ma_T \nabla_s T + \frac{\beta}{(1-\beta)Ca}\nabla_s \Gamma + \frac{\sigma}{Ca}(\nabla \cdot \hat{\mathbf{n}})\hat{\mathbf{n}},\end{aligned}\right\} \quad (16)$$

where, $r_s = 1 + Ca g^{(Ca)}(\theta,\varphi) + Ca^2 g^{(Ca^2)}(\theta,\varphi)$. $g^{(Ca)}$ and $g^{(Ca^2)}$ are $O(Ca)$ and $O(Ca^2)$ correction to the shape of the droplet. The stress balance condition in Eq. (16) is obtained by substituting the non-dimensional form of the equation of state, given by

$$\sigma = 1 - \zeta Ma_T Ca T_s - \beta \Gamma, \quad (17)$$

in the dimensional stress balance Eq. (8). The surface tension written in the above equation, is based on the modified capillary number. $Ca$. We can thus write [62]

$$\sigma = \frac{\bar{\sigma}}{\bar{\sigma}_c (1-\beta)}. \quad (18)$$

Thus it is inevitable that the value of $\beta$ lies within 1 and 0. The dimensionless surfactant transport equation can be written as

$$Pe_s \nabla_s \cdot (\mathbf{u}_s \Gamma) = \nabla_s^2 \Gamma. \tag{19}$$

The surfactants on the droplet surface must also fulfill the mass conservation constraint, which is given by

$$\int_{\varphi=0}^{2\pi} \int_{\theta=0}^{\pi} \Gamma(\theta, \varphi) \sin\theta \, d\theta \, d\varphi = 4\pi. \tag{20}$$

The temperature field, as observed from the Eq. (13), is uncoupled from the flow field under the assumption of low thermal Péclet number. So temperature in either of the phases can be solved independently without solving for the flow field. However, this is not the situation when we try to obtain the solution for surfactant concentration as well as flow field. The flow field is coupled with both the temperature field as well as the surfactant distribution on the droplet surface, due to the presence of Marangoni stress. The surfactant concentration, on the other hand, is coupled to the flow field through the surfactant transport equation (Eq. (19)). Thus a direct analytical solution is not possible. We thus use an asymptotic method which is performed for two limiting cases: (i) Low surface Péclet number, $Pe_s \ll 1$, that signifies that the dominant mode of surfactant transport is surface diffusion and (ii) high surface Péclet number, $Pe_s \to \infty$, which indicates that surface advection is the main mode of surfactant transport along the droplet surface.

### III. ASYMPTOTIC SOLUTION

In this section we provide a brief description of the methodology used to solve the temperature and the flow field along with the surfactant distribution as well as cite some of the important results obtained from the analysis. For a detailed discussion of the asymptotic approach, one can refer to the supplementary material provided.

#### A. Solution for $Pe_s \ll 1$

As only a small deformation of the droplet is assumed, the order of magnitude of the surface Péclet number is taken to be the same as that of the capillary number, that is $Pe_s \sim Ca$. This can be mathematically expressed as

$$Pe_s = \kappa Ca, \tag{21}$$

where, $\kappa = a\bar{\sigma}_{eq}(1-\beta)/\mu_e D_s$ is called the property parameter as it depends mainly of the material properties and has a finite magnitude. The droplet deformation is thus solely a function of capillary number, for a given value of $\kappa$ and $\beta$.

We, thus in our asymptotic analysis, choose $Ca$ as the perturbation parameter. The surfactant transport equation in this limit of low $Pe_s$ can be written in the following format

$$\kappa Ca \nabla_s \cdot (\mathbf{u}_s \Gamma) = \nabla_s^2 \Gamma. \tag{22}$$

As we will be using the regular perturbation method to solve for the temperature and flow field, we can expand any generic variable $(\psi)$ in a power series in terms of $Ca$ in the following manner

$$\psi = \psi^{(0)} + \psi^{(Ca)} Ca + O(Ca^2), \tag{23}$$

where, $\psi^{(0)}$ is the leading order term corresponding to no deformation of the droplet and $\psi^{(Ca)}$ is $O(Ca)$ correction to the quantity, $\psi$, due to deformation of the droplet. Other terms indicate even higher order corrections due to droplet deformation. The surfactant concentration on the other hand is represented in the following manner

$$\Gamma = 1 + \Gamma^{(0)} Ca + \Gamma^{(Ca)} Ca^2 + O(Ca^3), \tag{24}$$

where, $\Gamma^{(0)}$ and $\Gamma^{(Ca)}$ are $O(Ca)$ and $O(Ca^2)$ correction to the surfactant concentration due to droplet deformation. The leading order contribution to the surfactant concentration $(\Gamma = 1)$ signifies the scenario when the surfactants are uniformly distributed without any droplet deformation $(Ca = 0)$.

Firstly, with the help of Eq. (23), we derive the leading order and $O(Ca)$ boundary conditions for the temperature field. As the leading order governing equations and boundary conditions for temperature are not coupled with the flow field equations, the temperature field can be solved independently. The temperature field as obtained from the leading order solution is given by

$$\left. \begin{array}{l} T_i^{(0)} = \left( \dfrac{3\zeta}{2+\delta} \right) r \cos\theta \\[2mm] T_e^{(0)} = \zeta \left\{ 1 + \dfrac{1}{r^3} \left( \dfrac{1-\delta}{2+\delta} \right) \right\} r \cos\theta \end{array} \right\} \tag{25}$$

Secondly, we obtain the leading order flow field boundary conditions which along with the surfactant transport equation of the same order are solved for the velocity and the pressure field as well as the surfactant concentration. The leading order droplet migration velocity, which is obtained by the application of force-free condition is given below

$$U_z^{(0)} = \begin{bmatrix} \dfrac{(3R^2-2)\kappa\beta + 3(3\lambda R^2 - 2\lambda + 2R^2)(1-\beta)}{3R^2\{\kappa\beta + (3\lambda+2)(1-\beta)\}} - \dfrac{e^2}{R^2} \\ +2\zeta Ma_T \dfrac{(1-\beta)}{(\delta+2)\{\kappa\beta + (3\lambda+2)(1-\beta)\}} \end{bmatrix},$$

$$U_x^{(0)} = U_y^{(0)} = 0,$$
(26)

where, $U_z^{(0)}$ is the axial velocity of the droplet, whereas $U_x^{(0)}, U_y^{(0)}$ are the cross stream migration velocity components. From the above expression we can see that in the absence of droplet deformation, there is no cross stream migration velocity of the droplet. In the above expression for $U_z^{(0)}$, the second term on the right hand side of Eq. (26) shows the effect of the eccentric position of the droplet, whereas the last term shows the role of temperature distribution along the droplet surface on its migration in the flow field. The leading order surfactant concentration which satisfies the surfactant transport equation is given by

$$\Gamma^{(0)} = \Gamma_{1,0}^{(0)} P_{1,0} + \Gamma_{3,0}^{(0)} P_{3,0} + \Gamma_{2,1}^{(0)} \cos\varphi P_{2,1},$$

where,

$$\Gamma_{1,0}^{(0)} = \frac{1-\beta}{\{(3\lambda+2-\kappa)\beta - 3\lambda - 2\}} \left( \frac{3\kappa Ma_T}{\delta+2} + \frac{2}{R^2} \right)$$
(27)

$$\Gamma_{3,0}^{(0)} = -\frac{7\kappa}{6R^2} \left\{ \frac{1-\beta}{(7\lambda+7-\kappa)\beta - 7\lambda - 7} \right\}$$

$$\Gamma_{2,1}^{(0)} = \frac{5e\kappa}{3R^2} \left\{ \frac{1-\beta}{(5\lambda+5-\kappa)\beta - 5\lambda - 5} \right\}$$

Thirdly, with the leading order solution at hand, we find out the $O(Ca)$ correction to droplet shape with the help of the normal stress boundary condition. The normal stress boundary condition obtained from the stress balance condition is derived on the deformed surface of the droplet, $r_s = 1 + Ca g^{(Ca)} + Ca^2 g^{(Ca^2)}$. The $O(Ca)$ correction to the droplet shape, $g^{(Ca)}$ is given by

$$g^{(Ca)} = L_{3,0}^{(Ca)} P_{3,0} + L_{2,1} \cos\varphi P_{2,1},$$

where,

$$L_{2,1}^{(Ca)} = -\frac{5e}{12R^2} \left\{ \frac{4\kappa\beta + (16+19\lambda)(1-\beta)}{\kappa\beta + 5(1+\lambda)(1-\beta)} \right\},$$
(28)

$$L_{3,0}^{(Ca)} = \frac{7}{60R^2} \left\{ \frac{5\kappa\beta + 3(11\lambda+10)(1-\beta)}{\kappa\beta + 7(1+\lambda)(1-\beta)} \right\}.$$

It can be inferred from the above expression that the $O(Ca)$ shape deformation of the droplet is independent of the interfacial temperature gradient or the thermal Marangoni stress $(Ma_T)$. However the surfactant distribution is altered as a result of the thermally induced Marangoni stress, which can be seen from Eq. (27). This variation in interfacial surfactant concentration, which is coupled with the flow field through the surfactant transport equation [Eq. (22)], alters the fluid flow and hence the migration velocity of the droplet. With the leading order solution as well as $O(Ca)$ deformation at hand we move forward towards determining the $O(Ca)$ temperature field $\left(T_i^{(Ca)}, T_e^{(Ca)}\right)$, flow field $\left(\mathbf{u}_{i,e}^{(Ca)}, p_{i,e}^{(Ca)}\right)$ and the surfactant concentration $\left(\Gamma^{(Ca)}\right)$.

So, fourthly, the $O(Ca)$ temperature field is obtained by solving the relevant boundary conditions at the deformed surface of the droplet, with the help of Eq. (28). The expression for the $O(Ca)$ temperature field is provided below

$$\left.\begin{aligned} T_i^{(Ca)} &= \sum_{n=0}^{4}\sum_{m=0}^{n}\left[a_{n,m}^{(Ca)}r^n\cos(m\varphi)+\hat{a}_{n,m}^{(Ca)}r^n\sin(m\varphi)\right]P_{n,m}(\cos\theta), \\ T_e^{(Ca)} &= \sum_{n=0}^{4}\sum_{m=0}^{n}\left[b_{-n-1,m}^{(Ca)}r^{-n-1}\cos(m\varphi)+\hat{b}_{-n-1,m}^{(Ca)}r^{-n-1}\sin(m\varphi)\right]P_{n,m}(\cos\theta), \end{aligned}\right\} \tag{29}$$

where, the constant coefficients are given in Appendix A.

Fifthly, the $O(Ca)$ flow field is solved. Towards this, we first solve all the $O(Ca)$ flow field boundary conditions along with the $O(Ca)$ surfactant transport equation simultaneously. On obtaining the velocity field we make use of the zero net drag condition, to obtain the droplet migration velocity, both axial and cross-stream. Their expressions are provided below

$$\left.\begin{aligned} U_z^{(Ca)} &= 0, \ U_y^{(Ca)} = 0, \\ U_x^{(Ca)} &= \frac{e}{c_3(\beta,\kappa,\lambda,\delta)}\left[c_1(\beta,\kappa,\lambda,\delta)+\zeta Ma_T c_2(\beta,\kappa,\lambda,\delta)\right], \end{aligned}\right\} \tag{30}$$

where the constants $c_1$, $c_2$ and $c_3$ in the above expression is given in Appendix B. The above expression of the cross-stream migration of the droplet shows the effect of shape deformation of the droplet. As can be seen from the above expression for cross-stream migration velocity, the thermal Marangoni stress explicitly has an effect on the cross stream migration velocity of the droplet even though a constant temperature gradient is applied in the axial direction. This is quite non-intuitive. Also as seen, there is no axial velocity present at any higher orders of perturbation.

Finally, we obtain the $O(Ca)$ surfactant concentration by simultaneously solving the boundary conditions for flow field and the surfactant transport equation. The expression for $O(Ca)$ surfactant concentration is given by

$$\Gamma^{(Ca)} = \begin{bmatrix} \Gamma^{(Ca)}_{0,0} + \Gamma^{(Ca)}_{1,1}\cos\varphi P_{1,1} + \Gamma^{(Ca)}_{2,0} P_{2,0} + \Gamma^{(Ca)}_{2,2}\cos(2\varphi) P_{2,2} \\ +\Gamma^{(Ca)}_{3,1}\cos\varphi P_{3,1} + \Gamma^{(Ca)}_{4,0} P_{4,0} + \Gamma^{(Ca)}_{4,2}\cos(2\varphi) P_{4,2} \end{bmatrix}, \quad (31)$$

where $\Gamma^{(Ca)}_{0,0}$ is obtained from the relation for mass conservation of surfactants as given in Eq. (20). The detailed expressions of the constant coefficients in the above equation are extremely lengthy and hence not presented in the paper. However, if requested, we are ready to provide the detailed expression of the different coefficients involved in Eq. (31).

## B. Solution for $Pe_s \gg 1$

Unlike the limiting case of low Péclet number, we have $Pe_s^{-1} \sim Ca$ for the case of high surface Péclet number. The main difference from the previous case lies in the surfactant transport equation, which for the present limiting case becomes [56]

$$\nabla_s \cdot (\mathbf{u}_s \Gamma) = 0. \quad (32)$$

All the quantities in this case is expanded in a power series with respect to $Ca$ as was done in the previous case. A similar approach is followed for this limiting case too. The temperature field for the leading order of perturbation is the same as was obtained for the case of $Pe_s \ll 1$. The flow field for leading order is next solved simultaneously with the surfactant concentration. To obtain the droplet migration velocity we again apply the force-free condition on the droplet. The leading order droplet migration velocity in the present limiting case is given below

$$\left. \begin{array}{l} U_z^{(0)} = \left(1 - \dfrac{2}{3R^2}\right) - \dfrac{e^2}{R^2}, \\ U_x^{(0)} = 0,\ U_y^{(0)} = 0, \end{array} \right\} \quad (33)$$

The leading order surfactant concentration as obtained from the surfactant transport equation is as follows

$$\Gamma^{(0)} = \Gamma^{(0)}_{1,0} P_{1,0} + \Gamma^{(0)}_{3,0} P_{3,0} + \Gamma^{(0)}_{2,1} \cos\varphi P_{2,1}$$

where,

$$\Gamma^{(0)}_{1,0} = -\left(1 - \frac{1}{\beta}\right)\left(\frac{3Ma_T}{\delta + 2} + \frac{2}{R^2}\right), \quad (34)$$

$$\Gamma^{(0)}_{3,0} = -\frac{7}{6R^2}\left(1 - \frac{1}{\beta}\right),$$

$$\Gamma^{(0)}_{2,1} = -\frac{5e}{3R^2}\left(1 - \frac{1}{\beta}\right),$$

The $O(Ca)$ correction to the droplet shape, $g^{(Ca)}$, as obtained from the normal stress boundary condition, is provided below

$$g^{(Ca)} = L^{(Ca)}_{2,1} \cos\varphi P_{2,1} + L^{(Ca)}_{3,0} P_{3,0},$$

where, (35)

$$L_{2,1} = -\frac{5}{3}\frac{e}{R^2}, \quad L_{3,0} = \frac{7}{12R^2}$$

We next proceed further to calculate the $O(Ca)$ temperature field. On solving the boundary conditions for temperature field at the deformed droplet surface $\left(r_s = 1 + g^{(Ca)} Ca\right)$, we obtain the temperature distribution inside and outside the droplet as

$$\left.\begin{array}{l} T_i^{(Ca)} = \sum_{n=0}^{4}\sum_{m=0}^{n}\left[ p_{n,m}^{(Ca)} r^n \cos(m\varphi) + \hat{p}_{n,m}^{(Ca)} r^n \sin(m\varphi) \right] P_{n,m}(\cos\theta), \\ T_e^{(Ca)} = \sum_{n=0}^{4}\sum_{m=0}^{n}\left[ q_{-n-1,m}^{(Ca)} r^{-n-1} \cos(m\varphi) + \hat{q}_{-n-1,m}^{(Ca)} r^{-n-1} \sin(m\varphi) \right] P_{n,m}(\cos\theta), \end{array}\right\} \quad (36)$$

where the constant coefficients are provided in Appendix C.

The $O(Ca)$ velocity and pressure field is next obtained by simultaneously solving the boundary conditions for flow field and surfactant transport along the deformed interface. The axial and cross-stream droplet migration velocity, thus found out by using the force-free condition, is given by

$$\left.\begin{array}{l} U_z^{(Ca)} = 0, \ U_y^{(Ca)} = 0, \\ U_x^{(Ca)} = -e\left[\frac{4 - 3\beta}{6R^4 \beta} + \zeta Ma_T\left\{\frac{1 - \beta}{R^2(\delta + 2)\beta}\right\}\right]. \end{array}\right\} \quad (37)$$

As can be seen from the above expression of cross-stream migration velocity, the thermal Marangoni stress has a direct effect on the cross stream migration of the droplet, although the temperature gradient is applied in the axial direction. It can be seen from Eq. (37) that an axially imposed temperature gradient is seen to have a significant effect on the magnitude as well as the direction of cross-stream migration velocity.

The $O(Ca)$ surfactant concentration is next expressed below, which is obtained from the surfactant transport equation for this order.

$$\Gamma^{(Ca)} = \begin{bmatrix} \Gamma_{0,0}^{(Ca)} + \Gamma_{1,1}^{(Ca)} \cos\varphi P_{1,1} + \Gamma_{2,0}^{(Ca)} P_{2,0} + \Gamma_{2,2}^{(Ca)} \cos(2\varphi) P_{2,2} \\ +\Gamma_{3,1}^{(Ca)} \cos\varphi P_{3,1} + \Gamma_{4,0}^{(Ca)} P_{4,0} + \Gamma_{4,2}^{(Ca)} \cos(2\varphi) P_{4,2} \end{bmatrix}, \quad (38)$$

where all the constant coefficients in the above equation are provided in Appendix D. The expression for $\Gamma_{0,0}^{(Ca)}$ can easily be obtained with the help of the mass conservation constraint as given in Eq. (20) and is given below

$$\Gamma_{0,0}^{(Ca)} = -\frac{6}{5}\Gamma_{2,1}^{(0)} L_{2,1}^{(Ca)} - \frac{2}{7}\Gamma_{3,0}^{(0)} L_{3,0}^{(Ca)} - \frac{6}{5}\left\{L_{2,1}^{(Ca)}\right\}^2 - \frac{5}{7}\left\{L_{3,0}^{(Ca)}\right\}^2, \quad (39)$$

## IV. RESULTS AND DISCUSSION

The prime result of our analysis is the droplet cross-stream migration velocity for the two limiting cases: (i) low surface Péclet number limit $(Pe_s \ll 1)$ and (ii) high surface Péclet number limit. We first begin our discussion with the low surface Péclet number limit $(Pe_s \to \infty)$.

### A. Low surface Péclet Limit

In this limiting condition the surfactant transport is primarily dominated by surface diffusion rather than interfacial advection of surfactants. In the present study, we focus on the variation of the steady state cross-stream migration velocity of the droplet $(U_x)$. The expression for the cross-stream migration velocity of the droplet in this limit is given below:

$$\mathbf{U}_x = Ca \frac{e}{c_3(\beta,\kappa,\lambda,\delta)} \left[c_1(\beta,\kappa,\lambda,\delta) + \zeta Ma_T c_2(\beta,\kappa,\lambda,\delta)\right] \mathbf{e}_x \quad (40)$$

It is seen from Eq. (30) that there is no effect of shape deformation on the axial migration velocity of the droplet. Hence no further investigation on the same is done. Towards providing a detailed analysis on the cross-stream migration of the droplet, we first show the variation of $U_x$ with the viscosity ratio, $\lambda,$ for different values of thermal Marangoni number $\left(Ma_T\right)$ in Fig. 2. We consider two different cases of applied thermal gradient, namely, a linearly increasing temperature field in the direction of the imposed flow $\left(\zeta=1\right)$ and a linearly decreasing temperature field in the same direction $\left(\zeta=-1\right)$.

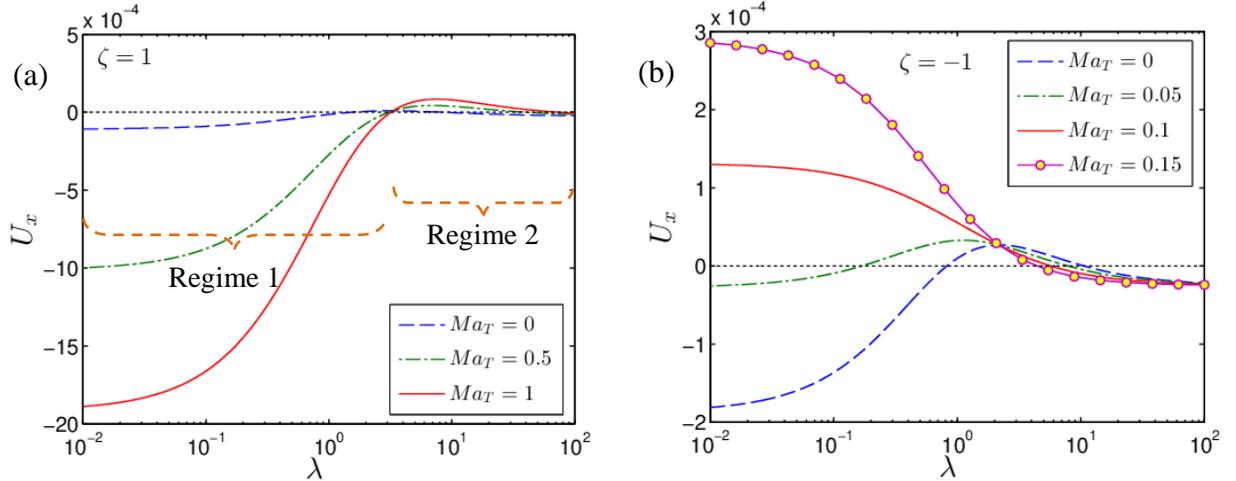

Fig. 2. Variation of cross-stream migration velocity of the droplet $\left(U_x\right)$ with $\lambda$ for different values of $Ma_T$. (a) Here the temperature increases in the direction of imposed Poiseuille flow $\left(\zeta=1\right)$ whereas in (b) the temperature decreases in the direction of the imposed flow $\left(\zeta=-1\right)$. The other parameters used are $\delta=1,\ \beta=0.5,\ \kappa=3,\ R=5,\ e=1$ and $Ca=0.1$.

### 1. Increase in temperature in the direction of the imposed Poiseuille flow $\left(\zeta=1\right)$

We first discuss the variation of the cross-stream migration velocity for the particular case of increasing temperature in the direction of bulk flow. It can be seen from Fig. 2(a) that for the case of a droplet suspended in an isothermal flow $\left(Ma_T=0\right)$, the cross-stream migration velocity decreases with increase in droplet viscosity (for a constant viscosity of the suspending medium). For a sufficiently high viscous droplet, the there is negligible cross-stream migration of the droplet. On the other hand, if an axial temperature gradient is externally imposed, the cross-stream migration velocity increases. For low values of $\lambda$, there is a significant effect of the applied temperature gradient on the cross-stream migration velocity of the droplet, but it gradually fades away as $\lambda$ is increased. This is because the jump in the tangential stress across the interface is significantly higher for the case of a low viscous droplet. For a non-isothermal system, the variation of the droplet cross-stream migration velocity with $\lambda$ follows the same trend as seen for $Ma_T=0$. It has also to be noted that there is an inflexion point present. That is, the droplet continues to migrate towards the centerline of flow as long as $\lambda<2.5$, depending on the other non-dimensional parameters which have been given in the caption of Fig. 2. This region has been named as 'regime 1' of the flow. Above $\lambda = 2.5$, the direction of cross-stream droplet migration changes, that is, it starts migrating away from the flow centerline although the cross-stream velocity still increases with increase in the axially applied temperature gradient $\left(\text{or } Ma_T\right)$. This region of flow where the direction of transverse migration of the droplet reverses is named as 'regime 2' of the flow. It should be noted that even though the temperature gradient is applied in the axial direction, the cross-stream migration of the droplet is affected. This is highly non-intuitive. On comparison with the study done by Das et al. [34], where the droplet was considered to be non-deformable, it can be seen that a deformable droplet may change its direction of cross-stream migration depending on the viscosity ratio of the system $(\lambda)$ for the special case of $\zeta = 1$. No such reversal in the direction of cross-stream migration was seen in the study done by Das et al.

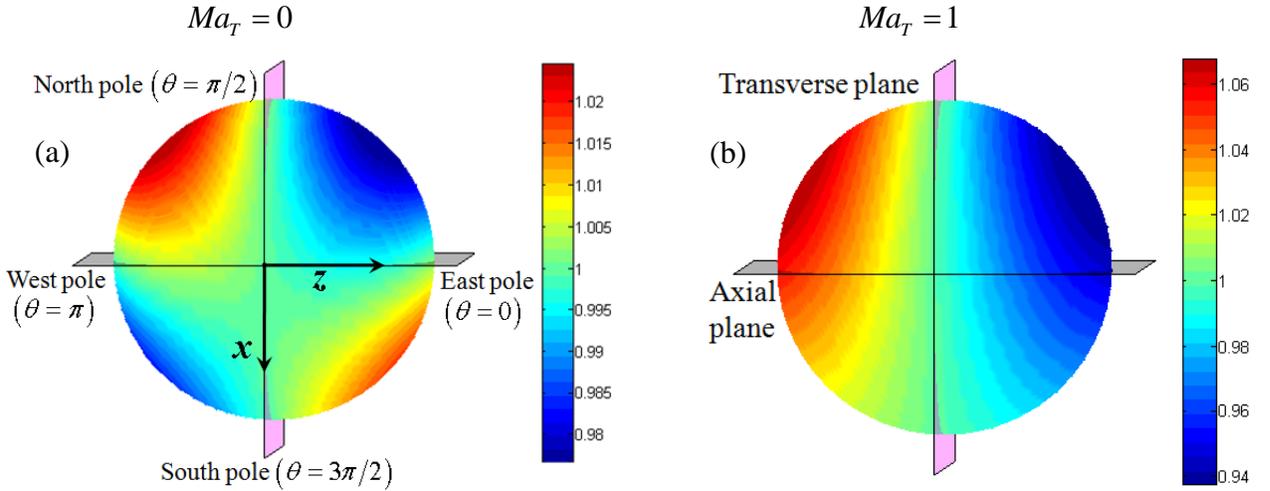

Fig. 3. Contour plot of the surfactant distribution $\left(\tilde{\Gamma}\right)$ on the droplet surface for two different cases, namely, (a) the droplet is suspended in an isothermal flow field $\left(Ma_T = 0\right)$ and (b) the droplet is suspended in a non-isothermal flow field with the temperature increasing in the direction of the imposed flow $\left(Ma_T = 1\right)$. The other parameter values for the above plot are taken as: $\delta = 1$, $\beta = 0.5$, $\lambda = 0.1$, $\kappa = 3$, $R = 5$, $e = 1$ and $Ca = 0.1$.

We now provide physical reasoning regarding the behavior of the cross-stream migration velocity of the droplet as seen above. Towards this, we first show the surfactant distribution along the droplet surface in the form of a contour plot in Fig. 3(a) and 3(b). Fig 3(a) shows the surfactant distribution for an isothermal flow field whereas Fig. 3(b) highlights the effect of a positive temperature gradient on the surfactant concentration and hence on the droplet dynamics. It can be seen from both Fig. 3(a) and 3(b) that the surfactant distribution is shown on an undeformed spherical droplet. This is done by projecting the surfactant distribution on a deformed droplet to a undeformed spherical surface of the droplet in the following form:

$\tilde{\Gamma} = \Gamma\left(r_s^2/\mathbf{n}\cdot\mathbf{r}\right)$ [31]. The sole reason for this transformation is to avoid the complexity of expressing the surface divergence vector on a deformed droplet. As our focus is primarily on the cross-stream migration of the droplet, we look into the surfactant distribution on either sides of the axial plane of the droplet. It can be seen from both Fig. 3(a) and 3(b) that there is a clear asymmetry in surfactant distribution across the axial plane. This is due to the eccentrically placed droplet in a Poiseuille flow as a result of which there exists unequal surface velocities along its northern and southern hemisphere. For instance consider the system shown in Fig. 1. As the droplet is placed below the centerline of flow, the upper hemisphere has a higher surface velocity as compared to the lower hemisphere, which is responsible for this asymmetry in surfactant distribution.

For the present limiting case of $\zeta = 1$, there is fluid flow from the east pole to the west pole along the droplet surface. Due to the presence of higher surface velocity in the northern hemisphere, there is a larger concentration of surfactants on the north-west part of the droplet as compared to the north-east portion (see Fig. 3(a)). The lower hemisphere, which has a lower surface velocity, has a higher concentration of surfactants on the south-east portion of the droplet. This asymmetry in surfactant distribution across the axial plane generates a gradient in the surface tension along any transverse plane of the droplet which in turn results in the creation of a Marangoni stress, responsible for the retardation of the cross-stream migration of the droplet [32,35]. The asymmetric distribution of surfactants and its transport along the droplet surface also significantly affects the normal stress balance and hence the associated droplet deformation [35,39]. Deformation of the droplet redistributes the surfactants along the droplet surface which alters the Marangoni stress and hence affects the droplet dynamics. However, the scenario gets interesting for the case of a non-isothermal fluid flow (see Fig. 3(b)). When a temperature gradient is applied in the direction of imposed flow, a thermal Marangoni stress is developed that opposes the Marangoni stress generated due to the non-uniform distribution of surfactants and forces the droplet to migrate towards the hotter region of the flow field. This increases the net convective transport of surfactants due to enhanced interfacial fluid flow as compared the case of an isothermal flow field. Hence an increase in the asymmetry of surfactant distribution along the droplet surface is expected. As can be seen from comparison of fig 3(b) and 3(a), there is a significant increase in the gradient in surfactant concentration $\left(\left|\Gamma_{max} - \Gamma_{min}\right|\right)$ when a constant temperature gradient is applied in the direction of bulk flow. In other words, there is an increase in asymmetry of the surfactant distribution as well. This results in a net increase in the surface tension gradient $\left(\left|\sigma_{max} - \sigma_{min}\right|\right)$ over the droplet surface which can be seen from Fig. 4(a) and Fig. 4(b). Both these figures show the variation of surface tension along two axial planes at two different transverse positions $(\theta = \pi/4, 3\pi/4)$ for a isothermal and a non-isothermal system, respectively. From Fig. 4(a) it is evitable that the surface tension is higher along the northern hemisphere whereas its magnitude is comparatively lower along the lower hemisphere. This generates a surfactant Marangoni stress that acts in a direction away from the flow centerline and hence opposes the imposed flow-driven cross-stream migration of the droplet towards the center.

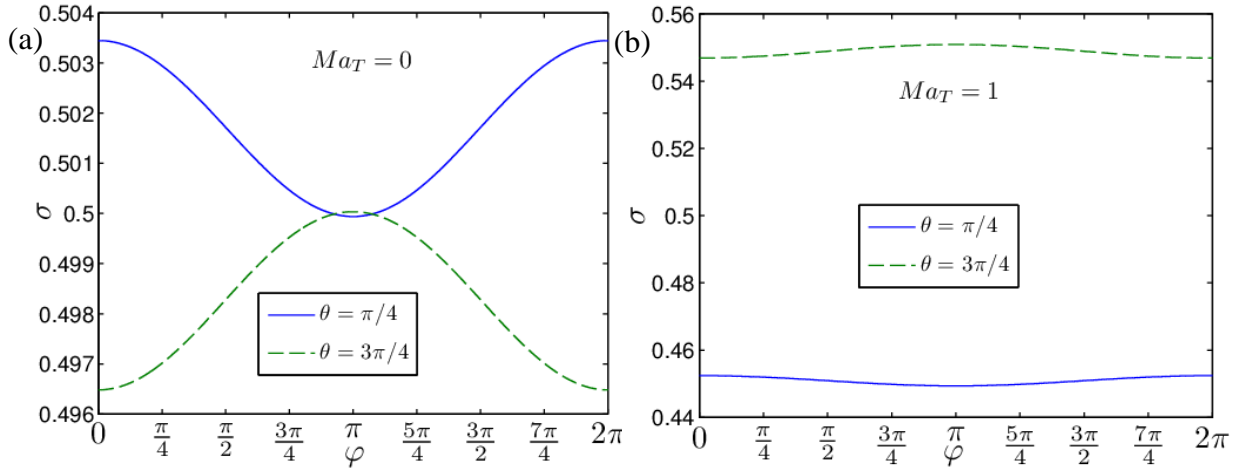

Fig. 4. Variation of surface tension along two different planes which are parallel to and are located on either sides of the axial plane, for two different cases, namely, (a) the droplet is suspended in an isothermal flow field $(Ma_T = 0)$ and (b) the droplet is suspended in a non-isothermal flow field with the temperature increasing in the direction of the imposed flow $(Ma_T = 1)$. The parameter values for the above plot are: $\delta = 1$, $\beta = 0.5$, $\lambda = 0.1$, $\kappa = 3$, $R = 5$, $e = 1$ and $Ca = 0.1$.

However, in Fig. 4(b), presence of a constant temperature gradient in the direction of the imposed flow results in a higher surface tension along the lower hemisphere, which indicates that the net Marangoni stress acts in a direction towards the centerline of flow, thus aiding droplet migration. Hence the cross-stream migration velocity increases. Further increase in the temperature gradient increases the net Marangoni stress, which in turn enhances the cross-stream velocity. This, in fact, was our observation from Fig. 2(a) in the regime 1. For 'regime 2', a high enough value of $\lambda$, alters the surfactant distribution along the droplet surface resulting a net change in the overall Marangoni stress, which now drives the droplet away from the flow centerline (Fig. 2(b)). In this regime too, increase in the temperature gradient is manifested by a rise in the magnitude of the cross-stream migration velocity of the droplet.

## 2. Decrease in temperature in the direction of the imposed Poiseuille flow $(\zeta = -1)$

We next discuss on the special case of a linearly decreasing temperature field in the direction of imposed bulk flow $(\zeta = -1)$. Fig. 2(b) shows the variation of cross-stream migration velocity with viscosity ratio, $\lambda$. In the present scenario too, the droplet initially, for low values of $\lambda$, migrates towards the centerline of flow, provided that the applied temperature gradient is sufficiently low $(\text{low } Ma_T)$. Unlike the previous case of $\zeta = 1$, the cross-stream migration velocity of the droplet in this case, reduces as the temperature gradient is increased. This decrease in the cross-stream migration velocity of the droplet continues till a point is reached where there is no more lateral migration. This is the critical point and the corresponding thermal Marangoni number is known as the critical thermal Marangoni number, $Ma_T^*$. Increase of $Ma_T$ beyond it critical value results in a reversal in the direction of the cross-stream migration of the droplet, which now migrates away from flow centerline. The expression for $Ma_T^*$ corresponding to zero cross-stream migration is can be obtained from Eq. (40) and is given below

$$Ma_T^* = \frac{c_1}{c_2}. \tag{41}$$

Thus any increase in the imposed temperature gradient, in this region, results in an increase in the magnitude of the cross-stream velocity of the droplet. This behavior can be seen for the case of low viscous droplets $(\lambda < 2)$. On the contrary for $\lambda > 2$, that is, for high viscous droplets, the behavior of the cross-stream migration of the droplet was noted to be just the opposite. Initially for the droplets with $\lambda \approx 2$, the cross-stream migration velocity reduces with increase in both $\lambda$ as well as $Ma_T$ and the droplet migrates away from the centerline of flow. On further increase in $\lambda$, the variation of cross-stream migration velocity, due to change in $Ma_T$, becomes negligible. Such high values of $\lambda$, results in droplet migration towards the flow centerline with a larger cross-stream migration velocity, irrespective of the value of $Ma_T$. Hence, foinsir sufficiently large $\lambda$, the droplet behaves as a particle and as seen from both Fig. 2(a) and 2(b), there is no effect of Marangoni stress on the droplet dynamics.

To obtain a better understanding, we now provide a physical insight to the scenario where the temperature gradient is applied opposite to the direction of the imposed flow $(\zeta = -1)$. Towards this, we show the distribution of surfactants on the droplet surface $\left[\tilde{\Gamma}(\theta,\varphi)\right]$ for $Ma_T < Ma_T^*$ and $Ma_T > Ma_T^*$ in a contour plot in Fig. 5(a) and 5(b), respectively. For the present scenario of temperature decreasing in the direction of the imposed fluid flow, the thermally-induced Marangoni stress drives the surfactants from the west pole to the east pole of the droplet, whereas the bulk Poiseuille flow forces the surfactants to migrate from the east to the west pole. Thus the surface fluid flow, induced by the imposed flow and the applied temperature gradient oppose each other. In addition to this, further surfactant redistribution takes place due to droplet deformation. Hence, depending on whether the Marangoni stress due to imposed flow or the applied temperature gradient dominates, the interfacial fluid flow may be from the west pole to the east pole or in the opposite direction. This net surface velocity along with the surfactant redistribution due to droplet deformation decides the direction as well as magnitude of the cross-stream migration velocity. For the case when the thermal Marangoni stress is less than its critical counterpart, the Marangoni convection due to the imposed Poiseuille flow dominates. Thus fluid flow takes place from the east to the west pole, and taking into account the nonuniformity in the surface velocity along the northern and the southern hemisphere of the droplet due to its eccentric positioning, the highest surfactant concentration is expected in the north-west region of the droplet surface, and the minimum surfactant concentration in the north-east region. This is exactly what is seen in Fig. 5(a). It can be seen that Fig. 5(a) is quite similar to that shown in Fig. 3(b). Hence the droplet, for the case of $Ma_T < Ma_T^*$, migrates towards the center line of flow. For the other case where $Ma_T > Ma_T^*$, the thermal Marangoni stress dominates and the direction of the surface velocity reverses. This results in a maximum surfactant concentration in the south-east region of the droplet, as shown in fig 5(b). On comparison with Fig. 5(a), the surfactant distribution is found to be just the opposite for this scenario. Hence the direction of the cross-stream migration alters and it starts migrating away from the flow centerline, as was shown in Fig. 2(b).

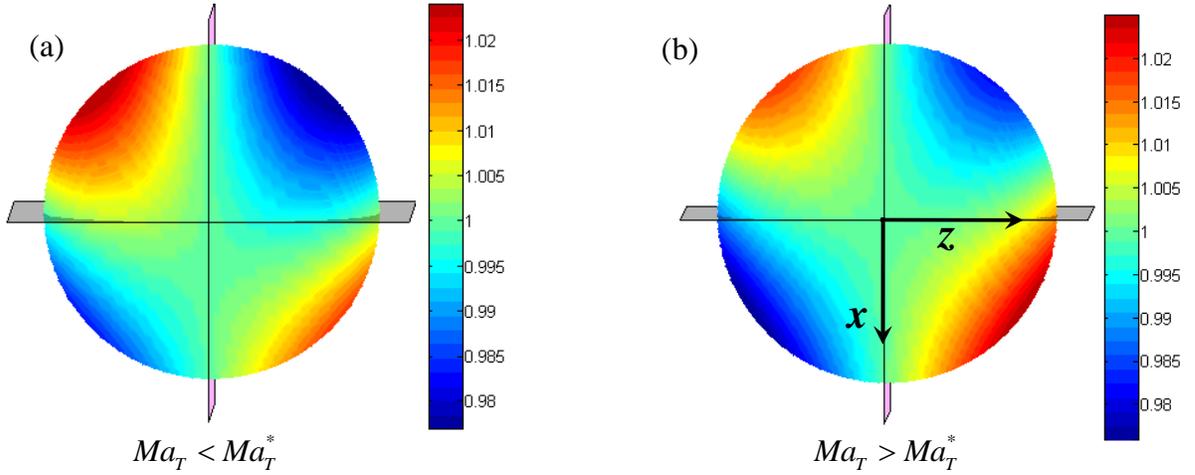

Fig. 5. Contour plot of the surfactant distribution $\left(\tilde{\Gamma}\right)$ on the droplet surface for the case when the temperature linearly decreases along the direction of the imposed flow $(\zeta = -1)$. The plot has been shown for two different situations, namely, (a) $Ma_T < Ma_T^*$ and (b) $Ma_T > Ma_T^*$. The parameter values for the above plot are: $\delta = 1$, $\beta = 0.5$, $\lambda = 0.1$, $\kappa = 3$, $R = 5$, $e = 1$ and $Ca = 0.1$.

This behavior can also be confirmed if we look into the variation of surface tension along the droplet surface corresponding to the above two cases. Fig. 6(a) and 6(b), shows the variation of $\sigma(\theta, \varphi)$ along two planes on either side of the axial plane at transverse positions, $\theta = \pi/4, 3\pi/4$. First of all, on comparison with Fig. 4(a), it can be seen that the net surface tension gradient, $|\sigma_{max} - \sigma_{min}|$, is higher for the case $Ma_T < Ma_T^*$. In addition, it is the upper droplet surface which has the higher surface tension as compared to the lower surface. Thus the net Marangoni stress developed acts away from the flow centerline. Since for this scenario the imposed Poiseuille flow dominates the interfacial fluid flow, the enhanced Marangoni stress due to increase in the surface tension gradient in comparison to the case of an isothermal flow $(Ma_T = 0)$, opposes the cross-stream migration of the droplet towards the flow centerline even more. As a consequence, the cross-stream migration velocity reduces with increase in $Ma_T$ as long as $Ma_T < Ma_T^*$ although the droplet migrates towards the axial plane. This can be seen in Fig. 2(b). For the special case when $Ma_T > Ma_T^*$, the thermally induced Marangoni stress dominates the interfacial fluid flow, and hence the net Marangoni stress, this time, succeeds in driving the droplet away from the flow center line. With increase in $Ma_T$, the increase in the net surface tension gradient increases, which indicates a net increase in the Marangoni stress. Thus further increase in $Ma_T$, enhances the cross-stream migration velocity of the droplet, which now migrate away from axial plane. However, for $\lambda > 2$ or droplet with higher viscosities, no particular critical point can be defined. The droplet, irrespective of the value of $Ma_T$, migrates away from the flow centerline. On further increase in $\lambda$, a point is reached, beyond which any increase in $Ma_T$ reduces the net cross-stream migration velocity. This continues till a certain value of $\lambda$ (see Fig. 2(b)). Further increase in $\lambda$, again alters the direction of droplet migration (towards the centerline) and finally the cross-stream migration velocity as $\lambda \to \infty$, which indicates the particle behavior of the droplet. This strange behavior in cross-stream migration of a highly viscous droplet is entirely due to variation in $\lambda$ and $Ma_T$ has little role to play in it.

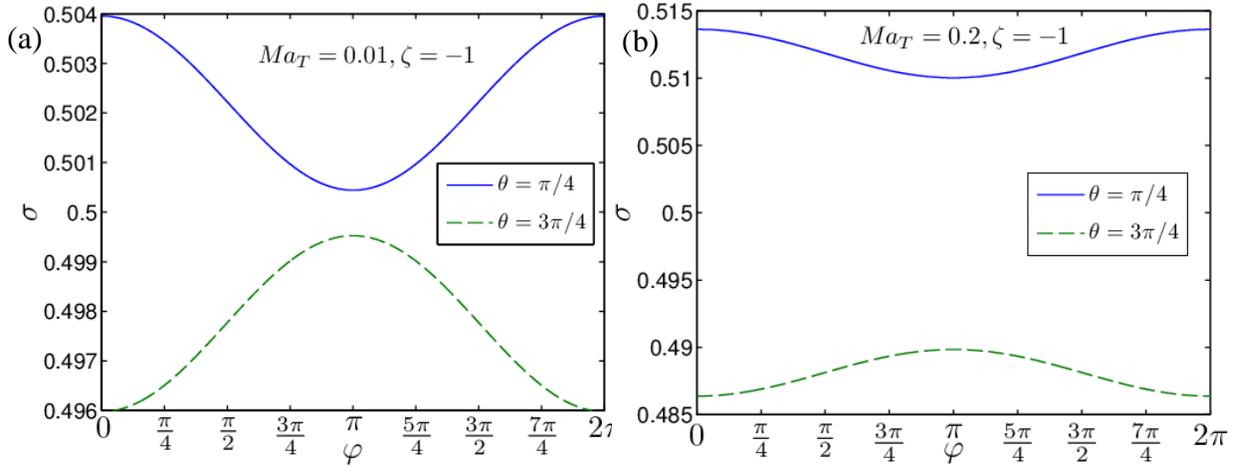

Fig. 6. Variation of surface tension along two different planes $(\theta = \pi/4, 3\pi/4)$ parallel to and located on either sides of the axial plane, for $\zeta = -1$ that is, the case when the temperature linearly decreases in the direction of the imposed flow. In fig. (a) $Ma_T < Ma_T^*$ and in fig. (b) $Ma_T > Ma_T^*$. The parameter values for the above plot are: $\delta = 1$, $\beta = 0.5$, $\lambda = 0.1$, $\kappa = 3$, $R = 5$, $e = 1$ and $Ca = 0.1$.

## 3. Cross-stream trajectory of the droplet as function of time

The transverse migration of the droplet as a function of time can be obtained by substituting $U_x = de/dt$ in the expression of the cross-stream migration velocity, given in Eq. (40). On integration and with the use of the initial condition $e(t=0) = e_0$ we get the expression for $e$ as

$$e = e_0 \exp\left(-\frac{t}{t_c}\right),$$

$$\text{where} \quad t_c = -\frac{c_3}{c_1 + \zeta Ma_T c_2}$$

(42)

Here $t_c$ is the characteristic time constant. It can be inferred from the above expression for the temporal variation of the lateral position of the droplet that a higher value of $t_c$ indicates a larger time for the droplet to reach its steady state position, which is the centerline of the flow field. Thus for the case when the temperature increases in the direction of imposed flow $(\zeta = 1)$, increase in $Ma_T$ results in a decrease in the magnitude of $t_c$ (as evitable from the expression of $t_c$).

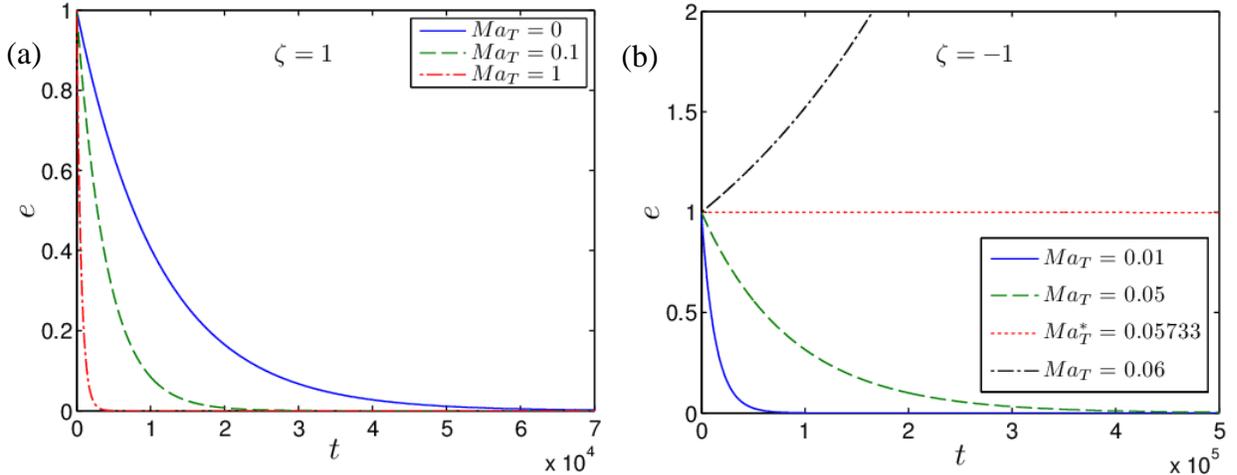

Fig. 7. Temporal variation of the transverse position of the droplet for different values of $Ma_T$, for the two different cases, (a) temperature increasing in the direction of imposed flow, $\zeta = 1$, and (b) temperature decreasing in the direction of imposed flow, $\zeta = -1$. The parameter values for the above plot are: $\delta = 1$, $\beta = 0.5$, $\lambda = 0.1$, $\kappa = 3$, $R = 5$, $e = 1$ and $Ca = 0.1$.

This variation in the transverse position of the droplet is confirmed from Fig. 7(a). Hence, as the applied temperature gradient in the direction of imposed flow increases, the time taken by the droplet to reach its steady state position reduces. The variation in the lateral position discussed here belongs to 'regime 1' of the cross-stream migration of the droplet, where the droplet always migrates towards the centerline of flow. This is quite non-intuitive as the applied temperature gradient in the axial direction alters the time taken by the droplet to reach its steady-state transverse position.

For the case when the temperature decreases in the direction of imposed fluid flow $(\zeta = -1)$, increase in $Ma_T$ increases the magnitude of $t_c$ (refer to the expression of $t_c$). As long as $Ma_T < Ma_T^*$, any increase in $Ma_T$ results is an increase in the time taken by the droplet to reach the flow centerline (steady state position). This is evitable from Fig. 7(b). When $Ma_T = Ma_T^*$, there is no lateral migration of the droplet (refer to Fig. 7(b)). It can also be seen from Fig. 7(b) that the droplet always migrates away from the channel centerline for $Ma_T > Ma_T^*$. However, for a highly viscous droplet as compared to the continuous phase, the droplet migrates towards the flow centerline irrespective of the value of $Ma_T$.

### B. High surface Péclet limit

Under this limit, the surfactant transport is along the interface is dominated by the surface convection. Hence in this limiting case, there is a higher fluid flow along the droplet surface as compared to that for the low surface Péclet number limit. The cross-stream migration velocity of the droplet for the limiting case of high $Pe_s$ is given by

$$\mathbf{U}_x = -eCa\left[\frac{1}{6R^4}\left(\frac{4-3\beta}{\beta}\right) + \frac{\zeta Ma_T}{R^2(\delta+2)}\left\{\frac{1-\beta}{\beta}\right\}\right]\mathbf{e}_x. \tag{43}$$

It can be inferred from the above expression that shape deformation play an important role in the cross-stream migration of the droplet. There is no presence of cross-stream migration velocity for the leading order solution. Since $\beta$ lies between 0 and 1, the above expression clearly indicates that the droplet always migrates towards the flow centerline when the temperature increases in the direction of the imposed flow $(\zeta = 1)$ which is unlike the case for the low Péclet number limit. However, for an applied temperature gradient in the opposite direction $(\zeta = -1)$, the droplet may either migrate towards or away from the flow centerline. In the absence of $\lambda$ in the above expression of cross-stream migration velocity, we show the variation of the same with $\beta$ for different values of $Ma_T$ in Fig. 8.

### 1. Increase in temperature in the direction of imposed flow $(\zeta = 1)$

Figure 8 shows the variation of $U_x$ as a function of $\beta$ for two separate cases. In Fig. 8(a) we show the variation for $\zeta = 1$ whereas in Fig. 8(b) the variation is shown for $\zeta = -1$. Each of the plots are shown for different values of $Ma_T$ to show the impact of the axially applied temperature gradient on the cross-stream migration velocity. The other parametric values are provided in the figure caption. We first analyze the case when the temperature increases in the direction of the imposed flow. It can be observed from Fig. 8(a) that increase in $\beta$ reduces the cross-stream migration velocity of the droplet irrespective of the applied temperature gradient. This is due to the fact that rise in $\beta$ actually increases the surfactant-induced Marangoni stress along the droplet interface that acts against the direction of the imposed flow and hence reduces the net cross-stream migration velocity of the droplet [35]. In the presence of an imposed axial temperature gradient, the cross-stream migration velocity increases in a similar manner as in the low Péclet limit. However, the magnitude of the cross-stream migration velocity is much larger as compared to the former limiting case due to surface convection dominated surfactant transport. It can also be seen from Fig. 8(a) that the impact of $Ma_T$ on the cross-stream migration velocity of the droplet reduces with increase in $\beta$. A higher value of $\beta$ results in an increased asymmetry in the surface tension across the axial plane and hence a larger surfactant-induced Marangoni stress, which nullifies the positive effect of the thermally-induced Marangoni stress.

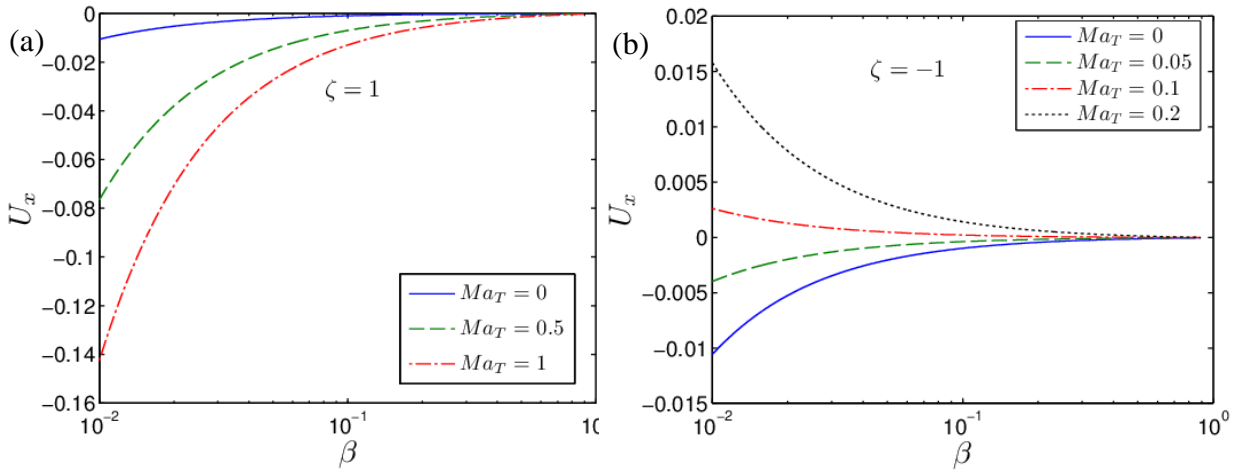

Fig. 8. Variation of cross-stream migration velocity $(U_x)$ with $\beta$ for different values of $Ma_T$. (a) Here the applied temperature gradient increases in the direction of imposed Poiseuille flow $(\zeta = 1)$. The values of $Ma_T$ used are $0, 0.5, 1$. (b) Here the applied temperature gradient decreases in a direction of the imposed flow $(\zeta = -1)$. The values of $Ma_T$ used in this Fig. are $0, 0.05, 0.1, 0.2$. The other parameters used in the above two figures are $\delta = 1$, $\beta = 0.5$, $\kappa = 3$, $R = 5$, $e = 1$ and $Ca = 0.1$.

To get a better physical understanding regarding the nature of variation of cross-stream migration velocity of the droplet due to change in $Ma_T$, we show a contour plot for the distribution of the surfactants along the surface of the droplet in Fig. 9. We again project the surfactant concentration on the deformed surface of the droplet to a spherical undeformed surface using the relationship $\tilde{\Gamma} = \Gamma\left(r_s^2/\mathbf{n}\cdot\mathbf{r}\right)$ [31] as was done in the other limiting scenario. It is to be noted from comparison between Fig. 9(a) and Fig. 3(a) that asymmetry in the surfactant distribution across the axial plane is higher for the limiting case of high Péclet number even for an isothermal flow field. This is, as explained before, due to enhanced convection driven surfactant transport. Similar is the case for a non-isothermal flow field (Fig. 9(b)) where too the magnitude of the asymmetry in surfactant distribution along the droplet surface $\left(|\Gamma_{max} - \Gamma_{min}|\right)$ is higher for the present limiting case. As the temperature increases in the direction of the bulk flow, there is an interfacial fluid flow from the east pole to the west pole of the droplet which together with the surface velocity due to the imposed flow results in the highest surfactant concentration along the north-west region of the droplet and lowest in the north-east region. Since the distribution pattern of the surfactants is similar to the limiting case of low Péclet number, the droplet in this scenario migrates towards the flow centerline. Increase in $\left(|\Gamma_{max} - \Gamma_{min}|\right)$ and hence in $\left(|\sigma_{max} - \sigma_{min}|\right)$ for a droplet suspended in a non-isothermal flow field results in an enhanced net Marangoni stress that drives the droplet towards the centerline of flow with a higher cross-stream migration velocity. This explains the fact that although the temperature gradient is applied axially, there is a significant increase in the cross-stream migration velocity of the droplet as well.

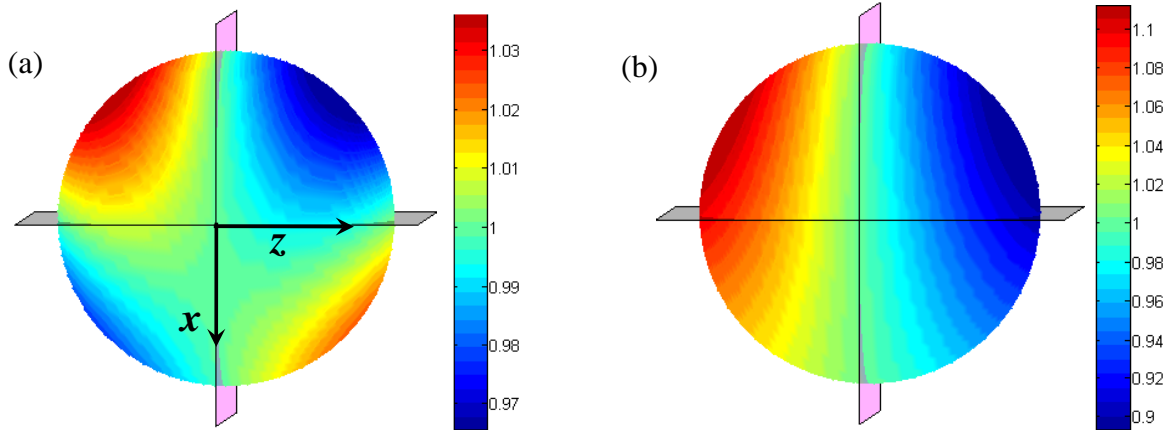

Fig. 9. Contour plot of the surfactant distribution $\left(\tilde{\Gamma}\right)$ on the droplet surface for two separate cases: (a) isothermal flow field $\left(Ma_T = 0\right)$ and (b) non-isothermal flow field with the temperature increasing in the direction of the imposed flow $\left(Ma_T = 1\right)$. The contour plot is shown for the limitng case of $Pe_s \to \infty$. The parameter values for the above plot are: $\delta = 1$, $\beta = 0.5$, $\lambda = 0.1$, $R = 5$, $e = 1$ and $Ca = 0.1$.

**2. Decrease in temperature in the direction of imposed fluid flow $\left(\zeta = -1\right)$**

We now look in the case where the temperature decreases linearly in the direction of the imposed flow. We thus refer to Fig. 2(b), where the variation of the cross-stream migration velocity is shown for this scenario. In this case too, irrespective of the magnitude of the applied temperature gradient (or $Ma_T$), increase in $\beta$ always results in a reduction of the cross-stream migration velocity. Similar to the previous limiting case for $Pe_s \ll 1$, we again define a critical Marangoni number, $Ma_T^*$, which denotes the critical point above which the droplet migrates away from the flow centerline and below which it moves towards the centerline of flow. At this critical point there is no cross-stream migration of the droplet. Keeping this in mind the expression of $Ma_T^*$ can be derived and expressed as

$$Ma_T^* = \frac{(\delta+2)}{6R^2}\left(3+\frac{1}{1-\beta}\right). \tag{44}$$

As long as $Ma_T < Ma_T^*$, any increase in the temperature gradient and hence $Ma_T$ results in a reduction in the magnitude of the cross-stream migration velocity. However, in the regime of $Ma_T > Ma_T^*$, the cross-stream migration velocity gradually increases with increase in $Ma_T$ and at the same time migrates away from the flow centerline. In this case too, the effect of $Ma_T$ on the cross-stream migration of the droplet reduces with increase in $\beta$.

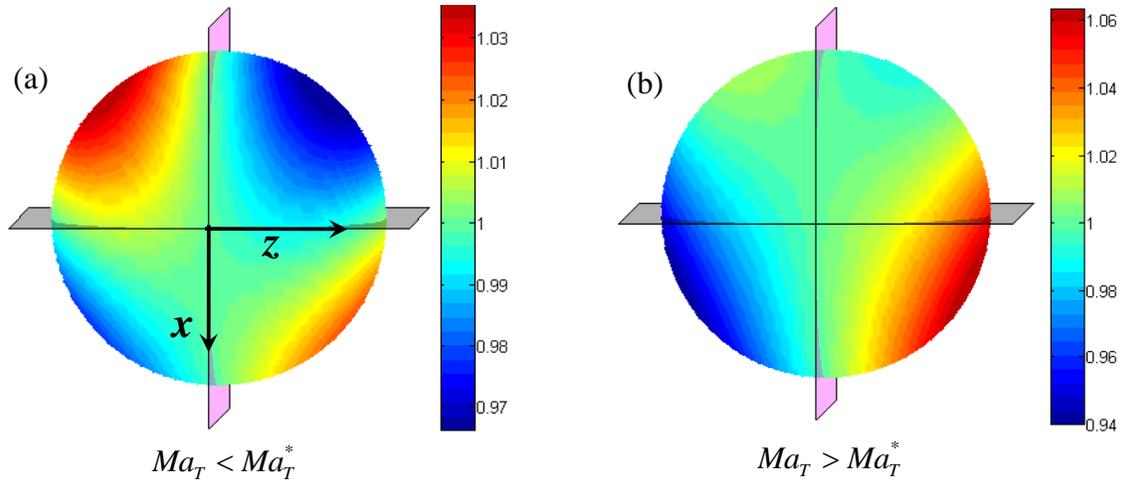

$Ma_T < Ma_T^*$          $Ma_T > Ma_T^*$

Fig. 10. Contour plot of the surfactant distribution $\left(\tilde{\Gamma}\right)$ on the droplet surface for two different scenarios: (a) $Ma_T = 0.01$ and (b) $Ma_T = 0.5$. The above figures are shown for the case of linearly decreasing temperature in the direction of imposed flow and for the limiting case of high surface Péclet number $(Pe_s \to \infty)$. The parameter values for the above plot are: $\delta = 1$, $\beta = 0.5$, $\lambda = 0.1$, $R = 5$, $e = 1$ and $Ca = 0.1$.

A better understanding on the same can be obtained if we look into the contour plots of the surfactant distribution along the droplet surface. Fig. 10(a) and 10(b) show the surfactant concentration along the droplet surface for the two special cases of $Ma_T < Ma_T^*$ and $Ma_T > Ma_T^*$ respectively. For the present scenario, the temperature-induced Marangoni stress acts opposite to the direction of the imposed flow. Hence the net surface velocity is dependent on which of the above factor is the dominant one. The net surface velocity alongside the shape deformation of the droplet decides the final surfactant distribution along the droplet interface. Depending on the distribution of the surfactants, the droplet may either migrate away from or towards the flow centerline. When $Ma_T < Ma_T^*$ (Fig. 10(a)), the imposed flow dominates over the thermocapillary effect and hence there is a net flow from the east pole to the west. Since the droplet is eccentrically located, the highest surfactant concentration is on the north-western region of the droplet while the minimum concentration is present in the north-eastern domain. This type of surfactant distribution, which was also obtained in Fig. 9(a), suggests that the droplet should migrate towards the flow centerline. On the other hand, when the thermocapillary effect dominates due to a high temperature gradient $\left(Ma_T > Ma_T^*\right)$, the fluid flow along the droplet surface reverses and the highest surfactant concentration is on the south-eastern region. Such a distribution of surfactants was previously obtained in Fig. 5(b). This suggests that the droplet migrates away from the centerline of flow. No separate attempt has been made to reproduce figures on the temporal variation of the steady state transverse position of the droplet in this present limiting case, as it yields results of similar nature.

### C. Comparison of our results with previously published work in the literature

In this section we compare the magnitude of cross-stream migration velocity obtained under both the limiting conditions, with the results of some of the previously published studies. The different values of the cross-stream migration velocity as obtained from the present study as well from other related work done by Chan et al. [20] and Das et. al. [34,35] are tabulated below

Table 1: Comparison of the magnitude of the cross-stream migration velocity with the results obtained from previously published works.

| Different Studies | Cross-stream velocity ($U_x$) | |
|---|---|---|
| | $Pe_s \ll 1$ | $Pe_s \gg 1$ |
| Cross-stream migration of a surfactant-free deformable droplet in an isothermal Poiseuille flow [20] | $1.205 \times 10^{-4}$ | |
| Cross-stream migration of a surfactant-laden non-deformable droplet in a non-isothermal Poiseuille flow [34] | $4.828 \times 10^{-4}$ | $10.057 \times 10^{-4}$ |
| Cross-stream migration of a surfactant-laden deformable droplet in an isothermal Poiseuille flow [35] | $0.68 \times 10^{-4}$ | $0.667 \times 10^{-4}$ |
| Present study | $6.317 \times 10^{-4}$ | $10.19 \times 10^{-4}$ |

As can be seen from table 1 that under the presence of the a constant axial temperature gradient as well as consideration of droplet deformation results in a significant rise in the magnitude of the cross-stream migration velocity of the droplet. The values of the different parameters used for this evaluation are $Ca = 0.05$, $Pe_s = 0.1$, $\delta = 1$, $R = 5$, $e = 1$, $\lambda = 0.1$, $\kappa = 1$, $\beta = 0.5$. For the high Péclet number limit we have used $Pe_s = 10$.

## V. CONCLUSIONS

The present study deals with the cross-stream migration of a surfactant-laden droplet suspended in a Poiseuille flow with a linearly increasing temperature gradient. The droplet is taken to be deformable; however, only small deviations from the spherical shape are assumed. The system under consideration is taken to neutrally buoyant and any presence of inertia in fluid flow is neglected. We use a asymptotic approach to solve the nonlinear system of governing equations and relevant boundary conditions under two different limiting cases, namely, surface-diffusion-dominated and surface-convection-dominated transport of surfactants. Since the system of governing equations and boundary conditions are all highly non-linear and coupled due to the consideration of droplet deformation and associated surfactant redistribution, a mere superposition of the results for the thermocapillary-driven and Poiseuille flow-driven droplet migration is not possible. We obtain the droplet migration velocity as well as the surfactant concentration along the droplet surface till O(*Ca*). The thermocapillary effect on droplet cross-stream migration is analyzed for two specific cases: one in which the temperature increases linearly in the direction of the imposed flow $(\zeta = 1)$ and the other where the direction of the applied temperature gradient is reversed $(\zeta = -1)$. After a thorough analysis of the droplet migration characteristics, some of the important findings established are stated below

- For the limiting case of low surface Péclet number, the droplet, in general, always migrates towards the centerline of flow. It is seen that increase in the axial temperature gradient results in increase in the cross stream migration velocity of the droplet, provided the temperature increases in the direction of the imposed flow. For high viscosity ratios, the direction the droplet cross-stream migration reverses depending on the value of $Ma_T$.

- When the temperature decreases in the direction of the imposed Poiseuille flow, the droplet may migrate towards or away from the flow centerline depending on the magnitude of the applied temperature gradient as well as the droplet viscosity. For a low viscous droplet, the droplet migrates towards the flow centerline and the magnitude of the cross-stream migration velocity reduces with increase in $Ma_T$ till $Ma_T < Ma_T^*$. However, at the critical point $Ma_T = Ma_T^*$, there is no cross-stream migration. Beyond this critical value $\left(Ma_T^*\right)$ any further increase in $Ma_T$ results in an increase in the magnitude of the cross-stream migration velocity. The droplet now migrates away from the flow centerline. A highly viscous droplet, on the other hand, always migrates towards the centerline of flow.

- For the limiting case of high surface Péclet number limit, the magnitude of the cross-stream migration velocity is always higher as compared to the limiting case of low surface Péclet number. The nature of variation of the steady-state cross-stream velocity with $Ma_T$ is the same as that for the low Péclet number limit, but it is independent of the droplet viscosity.

**Supplementary material**

See the supplementary material for details regarding the governing equations, the boundary conditions and the asymptotic approach adopted in this study to obtain the temperature and the flow field.

**Acknowledgement**

Valuable inputs from Mr. Shubhadeep Mandal are gratefully acknowledged.

**Appendix A: Constant coefficients present in the expression for $O(Ca)$ temperature field in Eq. (29)**

The expressions of the constant coefficients present in the expression for $O(Ca)$ temperature field is given by

$$a_{1,1}^{(Ca)} = \frac{9e(1-\delta)}{R^2(2+\delta)^2} \left\{ \frac{\left(4 + \frac{19}{4}\lambda - \kappa\right)\beta - \frac{19}{4}\lambda - 4}{(-\kappa + 5 + 5\lambda)\beta - 5\lambda - 5} \right\},$$

$$b_{1,1}^{(Ca)} = \frac{3e(5+\delta)(1-\delta)}{R^2(2+\delta)^2} \left\{ \frac{\left(-4 - \frac{19}{4}\lambda + \kappa\right)\beta + \frac{19}{4}\lambda + 4}{(\kappa - 5 - 5\lambda)\beta + 5\lambda + 5} \right\},$$

(A1)

$$a_{2,0}^{(Ca)} = \frac{9(1-\delta)}{4R^2(2+\delta)(3+2\delta)} \left\{ \frac{\left(6+\frac{33}{5}\lambda-\kappa\right)\beta-\frac{33}{5}\lambda-6}{(\kappa-7-7\lambda)\beta+7+7\lambda} \right\},$$

$$b_{2,0}^{(Ca)} = \frac{3(1-\delta)(\delta+3)}{2R^2(2+\delta)(3+2\delta)} \left\{ \frac{\left(6+\frac{33}{5}\lambda-\kappa\right)\beta-\frac{33}{5}\lambda-6}{(\kappa-7-7\lambda)\beta+7+7\lambda} \right\},$$

(A2)

$$a_{3,1}^{(Ca)} = \frac{8e(1-\delta)}{R^2(4+3\delta)(2+\delta)} \left\{ \frac{\left(4+\frac{19}{4}\lambda-\kappa\right)\beta-\frac{19}{4}\lambda-4}{(\kappa-5-5\lambda)\beta+5\lambda+5} \right\},$$

$$b_{3,1}^{(Ca)} = \frac{6e\delta(1-\delta)}{R^2(4+3\delta)(2+\delta)} \left\{ \frac{\left(-4-\frac{19}{4}\lambda+\kappa\right)\beta+\frac{19}{4}\lambda+4}{(\kappa-5-5\lambda)\beta+5\lambda+5} \right\},$$

(A3)

$$a_{4,0}^{(Ca)} = \frac{6(1-\delta)}{R^2(5+4\delta)(2+\delta)} \left\{ \frac{\left(-6-\frac{33}{5}\lambda+\kappa\right)\beta+\frac{33}{5}\lambda+6}{(\kappa-7-7\lambda)\beta+7+7\lambda} \right\},$$

$$b_{4,0}^{(Ca)} = \frac{(1-\delta)(1-4\delta)}{R^2(5+4\delta)(2+\delta)} \left\{ \frac{\left(-6-\frac{33}{5}\lambda+\kappa\right)\beta+\frac{33}{5}\lambda+6}{(\kappa-7-7\lambda)\beta+7+7\lambda} \right\}.$$

(A4)

All the other coefficients present in the expression of $O(Ca)$ temperature field are zero.

## Appendix B: Constant coefficients present in the expression of $O(Ca)$ cross stream migration velocity in Eq. (30)

The constant coefficients present in the expression of cross stream droplet migration velocity are given below

$$c_1 = 2(\delta+2)^2 \begin{Bmatrix} 5\beta^3(3\beta-4)\kappa^4 - \beta^2(\beta-1)(152\beta\lambda+115\beta-240\lambda-190)\kappa^3 \\ +\beta(\beta-1)^2(142\beta\lambda^2+943\beta\lambda-240\beta-1090\lambda-700\lambda^2-400)\kappa^2 \\ +\beta(-4339\lambda^2+4220+1512\lambda^3-593\lambda)(\beta-1)^3\kappa \\ -15(99\lambda^4-1258\lambda^2-1017\lambda^3+486\lambda+640)(\beta-1)^4 \end{Bmatrix}$$

$$c_2 = 15R^2(\beta-1)\begin{pmatrix}\beta\kappa\\-7(1+\lambda)(\beta-1)\end{pmatrix}\begin{Bmatrix}4\beta^2(\delta+2)\kappa^3 \\ -4\beta\begin{pmatrix}-10\lambda+5\beta\lambda\delta-12bt\delta\\+12\beta-6-3\delta-5\lambda\delta+10\beta\lambda\end{pmatrix}\kappa^2 \\ -\beta(-16+381\lambda\delta+424\delta-354\lambda)(\beta-1)\kappa \\ +2(\beta-1)^2(16+19\lambda)(-30\lambda+12\lambda\delta+10+23\delta)\end{Bmatrix} \quad (B1)$$

$$c_3 = 60(\delta+2)^2(\beta\kappa-5\beta-5\beta\lambda+5+5\lambda)(\beta\kappa-7\beta-7\beta\lambda+7+7\lambda)(\beta\kappa-2\beta-3\beta\lambda+3\lambda+2)^2 R^4$$

**Appendix C: Constant coefficients present in the expression of $O(Ca)$ temperature field as shown in Eq. (36) for the limiting case of high surface Péclet number**

The expression for the constant coefficients present in the $O(Ca)$ temperature field are given below. All the other coefficients other than the ones presented below are zero.

$$\left.\begin{aligned}
p_{1,1}^{(Ca)} &= \frac{9e}{R^2}\left\{\frac{1-\delta}{(\delta+2)^2}\right\}, \quad q_{1,1}^{(Ca)} = \frac{3e}{R^2}\left\{\frac{(1-\delta)(\delta+5)}{(\delta+2)^2}\right\}, \\
p_{2,0}^{(Ca)} &= -\frac{9}{4R^2}\left\{\frac{1-\delta}{(\delta+2)(3+2\delta)}\right\}, \quad q_{2,0}^{(Ca)} = \frac{3}{2R^2}\left\{\frac{\delta^2+2\delta-3}{(\delta+2)(3+2\delta)}\right\}, \\
p_{3,1}^{(Ca)} &= -\frac{8e}{R^2}\left\{\frac{1-\delta}{(\delta+2)(4+3\delta)}\right\}, \quad q_{3,1}^{(Ca)} = \frac{6e}{R^2}\left\{\frac{\delta(1-\delta)}{(\delta+2)(4+3\delta)}\right\} \\
p_{4,0}^{(Ca)} &= \frac{6}{R^2}\left\{\frac{1-\delta}{(2+\delta)(5+4\delta)}\right\}, \quad q_{4,0}^{(Ca)} = \frac{1-5\delta+4\delta^2}{R^2(2+\delta)(5+4\delta)}.
\end{aligned}\right\} \quad (C1)$$

**Appendix D: Constant coefficients present in the expression of $O(Ca)$ surfactant concentration as shown in Eq. (38) for the limiting case of high surface Péclet number**

The expression of the constant coefficients in $O(Ca)$ surfactant concentration as given in Eq. (38) is written below

$$\Gamma_{1,1}^{(Ca)} = -\frac{9}{2} \frac{\left[\begin{array}{l}\left[\left(-\frac{7}{18}+\frac{2}{3}\lambda\right)\delta^2 + \left(\left(\zeta Ma_T R^2 + \frac{8}{3}\right)\lambda + \frac{17}{3}\zeta Ma_T R^2 - \frac{14}{9}\right)\delta \\ +\left(\frac{8}{3}+2\zeta Ma_T R^2\right)\lambda - \frac{2}{3}\zeta Ma_T R^2 - \frac{14}{9}\end{array}\right]\beta (-1+\beta)e}{R^4 (2+\delta)^2 \beta^2},$$

$$\Gamma_{2,0}^{(Ca)} = \frac{6(-1+\beta)\left(\begin{array}{l}\left(\left(-\frac{5}{18}+\left(\frac{10}{9}\lambda+\frac{5}{63}\right)e^2\right)\delta^2 + \left(\left(\frac{5}{18}+\frac{35}{9}\lambda\right)e^2 + \zeta Ma_T R^2 - \frac{35}{36}\right)\delta\right)\beta \\ +\left(\frac{5}{21}+\frac{10}{3}\lambda\right)e^2 - \frac{5}{6} - \frac{3}{8}\zeta Ma_T R^2 \\ -\frac{25}{18}(1+\lambda)(2+\delta)\left(\delta+\frac{3}{2}\right)e^2\end{array}\right)}{\beta^2 R^4 (2\delta^2 + 7\delta + 6)},$$

$$\Gamma_{2,2}^{(Ca)} = -\frac{5}{9} \frac{(-1+\beta)\left(\left(\frac{11}{14}+\lambda\right)\beta - \frac{5}{4} - \frac{5}{4}\lambda\right)e^2}{\beta^2 R^4}, \quad \Gamma_{4,2}^{(Ca)} = -\frac{25}{42}\frac{(-1+\beta)e^2}{\beta R^4},$$

$$\Gamma_{3,1}^{(Ca)} = \frac{10(-1+\beta)\left(\begin{array}{l}\left(\left(-\frac{35}{48}+\frac{7}{40}\lambda\right)\delta^2 + \left(\zeta Ma_T R^2 + \frac{7}{12}\lambda - \frac{175}{72}\right)\delta\right)\beta \\ -\frac{35}{18} - \frac{12}{5}\zeta Ma_T R^2 + \frac{7}{15}\lambda \\ -\frac{49}{240}(1+\lambda)(2+\delta)\left(\delta+\frac{4}{3}\right)\end{array}\right)e}{(3\delta^2 + 10\delta + 8)R^4 \beta^2}, \quad (D1)$$

$$\Gamma_{4,0}^{(Ca)} = -\frac{8(-1+\beta)\left(\begin{array}{l}\left(-\frac{13}{22}-\frac{25}{7}e^2\right)\delta^2 + \left(\zeta Ma_T R^2 - \frac{169}{88} - \frac{325}{28}e^2\right)\delta \\ -\frac{65}{44} - \frac{125}{14}e^2 - \frac{17}{8}\zeta Ma_T R^2\end{array}\right)}{(4\delta^2 + 13\delta + 10)R^4 \beta},$$